%% file: MAIN.tex
\documentclass{article}

\usepackage{PRIMEarxiv}
\usepackage{longtable}
\usepackage{graphicx}  
\usepackage{subfigure}
\usepackage[utf8]{inputenc} 
\usepackage[T1]{fontenc}    
\usepackage{hyperref}       
\usepackage{url}            
\usepackage{booktabs}       
\usepackage{amsfonts}       
\usepackage{nicefrac}       
\usepackage{microtype}      
\usepackage{lipsum}
\usepackage{fancyhdr}       
\usepackage{multirow}
\usepackage{xcolor}
\graphicspath{{media/}}     

\title{Usability of Privacy Controls in Top Health Websites
}
\author{
Ravin Gunawardena \\ University of New South Wales \\ \texttt{r.gunawardena@unsw.edu.adu} \And 
Yuemeng Yin \\ University of New South Wales \\ \texttt{yuemeng.yin@student.unsw.edu.au}  \And 
Yi Huang \\ University of Technology Sydney \\ \texttt{Yi.Huang-3@student.uts.edu.au}\And 
Rahat Masood \\ University of New South Wales \\ \texttt{rahat.masood@unsw.edu.au} \And 
Suranga Seneviratne \\ University of Sydney \\ \texttt{suranga.seneviratne@sydney.edu.au} \And 
Imran Razzak \\ University of New South Wales \\ \texttt{imran.razzak@unsw.edu.au} \And 
Nguyen Tran \\ University of Sydney \\ \texttt{nguyen.tran@sydney.edu.au} \And Aruna Seneviratne \\ University of New South Wales \\ \texttt{a.seneviratne@unsw.edu.au}
}

\begin{document}
\maketitle

\input{Sections/Abstract}


\input{Sections/Introduction}
\input{Sections/Related}
\input{Sections/Background}
\input{Sections/Methodology}

\input{Sections/Results}
\input{Sections/Discussion}

\input{Sections/Conclusion}


\bibliographystyle{unsrt}  
\bibliography{REF}

\end{document}

%% file: Sections/Abstract.tex
\begin{abstract}
With the increasing awareness and concerns around privacy, many service providers are offering various privacy controls to their users. Through these controls, users gain greater authority over the collection, utilization, and dissemination of their personal information by the services. However, these controls may be buried deep within menus or settings, making them difficult for a user to access. Additionally, the terminology used to describe privacy controls can sometimes be confusing or technical, further complicating the user's ability to understand and use them effectively. This is especially true for health websites, as users often share sensitive information about their health and well-being. While many privacy controls have been proposed to protect user data on these sites, existing research tends to focus on individual controls (e.g., privacy policies or cookie opt-outs) rather than providing a comprehensive overview of the privacy landscape. In addition, many studies concentrate on the technical aspects of privacy controls without considering the usability of these features from a user's perspective. 
This paper aims to fill the gaps in existing literature by providing an analysis of four privacy controls, namely privacy nudge, privacy notice, privacy policy, and privacy setting, and evaluating their usability on the top 100 most visited health websites. We define usability attributes for each privacy control in three website visit scenarios; the guest, registering, and log-in visits. These attributes include awareness, efficiency, comprehension, functionality, and choice. Then, we design a survey template 
based on these usability attributes and scenarios and collect data about privacy controls. Next, we analyse the availability and usability of each privacy control on health websites. Finally, the paper provides suggestions for improving the design of these privacy controls based on the data analysis results. Our analysis and recommendations can help website designers and policymakers improve the usability and effectiveness of privacy controls to protect users' privacy rights.
\end{abstract}

%% file: Sections/Introduction.tex
\section{Introduction}
\label{sec:introduction}

User privacy has been one of the most focused areas in the healthcare sector. With the growth of online resources for health-based services, the amount of people using these services has been increasing continuously. This raises concern for the security and privacy of the users utilizing web-based health services.
Service developers have developed various security and privacy tools and techniques to address some of these concerns including, preserving anonymity online and controlling what data is shared with third parties. 
For instance, privacy-preserving algorithms such as Differential Privacy have been designed to preserve user data to maintain user privacy. Similarly, various controls such as privacy nudges, cookie settings, and web browsers with extra privacy features such as Tor and Brave, have been developed to provide users with a higher degree of control over their online presence. In addition, regulations such as GDPR and CCPA enforce developers to follow responsible data collection practices. However, despite all these efforts, health websites act as continuous targets for hackers often leading to large-scale privacy breaches, exposing personal and intimate details of users. For example, recent data breach at Medibank compromised the health records of millions of Australian citizens, constituting a serious violation of their privacy rights. One major factor contributing to such breaches is the lack of user-friendly privacy controls on these websites. As a result, users either remain unaware of these controls or struggle to configure them correctly.



Quite recently, multiple research focused on the usability of privacy controls provided by online services. For example, data deletion and opt-out choices~\cite{habib2019empirical} and cookie management 
providers~\cite{toth2022dark,matte2020cookie}. 
These research predominantly considers only one or two aspects of privacy e.g., privacy policy or privacy nudges. Furthermore, existing works are yet to estimate the overall usability of privacy measures. In addition, only a limited amount of work investigated the usability of privacy measures on health websites that handle more private information.


To this end, we developed a holistic template to empirically analyse the usability of four different privacy controls (privacy nudges, privacy policies, privacy notices, and privacy settings). We perform this analysis on the top 100 most visited health websites~\cite{similarweb}. Since users are more likely to share private information with these websites, there is a high degree of concern. Our evaluation is performed under several scenarios where the user can either visit as a guest, a registering user, or an already registered user. First, we analyse the availability of the above privacy controls in these websites. Moreover, we consider the impact of the regulations on the availability of privacy controls by comparing all websites with the ones complying with regulations, especially
the most widely used GDPR and CCPA. Then, based on the data collected through our holistic template, we analyse and evaluate the usability of each privacy control. 

More specifically, we make the following contributions.
\begin{enumerate}
\item We have designed a holistic survey template that can be used to analyse and evaluate the usability of four different privacy controls. In this template we take into consideration various factors such as the location, display type, number of clicks, and content types across privacy controls. 

\item Based on the template, we empirically analysed the usability of the top 100 health websites under three scenarios; guest visit, registration, and login. We also compare how privacy controls change under each scenario and the impact they have on the usability.

\item We evaluated these websites based on health guidelines to determine whether they conform to these recommendations. 

\item Based on the findings of our survey, we propose several designs for privacy controls, as well as recommendations for integrating privacy policies and settings.
\end{enumerate}

%% file: Sections/Related.tex
\section{Related work}
\label{sec:relatedwork}

\subsection{Research on Privacy Controls}
GDPR and CCPA are two of the most commonly followed regulations by websites across the internet. Many researchers have analyzed the influence of these regulations alongside the regulatory compliance of websites that adhere to them. Degeling et al.~\cite{degeling2018we} evaluated the improvement to privacy controls of a website, such as the privacy policy and cookie consent with the implementation of GDPR. Their analysis showed that across 28 countries of the European Union, around 16\% of websites added privacy policies and cookie consent notices by the deadline provided by GDPR. Hils et al.~\cite{hils2020measuring} analyzed 161 million web pages and found that implementing GDPR and CCPA quadrupled the adoption of Consent Management Platforms (CMPs) from 2018 to 2020. However, researchers have found evidence of disparities between the displayed/user-selected privacy settings and the behaviour of websites. Matte et al.~\cite{matte2020cookie} crawled websites following IAB Europe’s Transparency and Consent Framework (TCF) and found that even when users opted out from consent, it was recorded as positive consent. Which violates the guidelines of the framework the website adheres to. Liu et al.~\cite{liu2022opted} demonstrated that many websites violated the data protection regulations by still collecting, processing and sharing user information after users have opted out. To this end, they proposed an auditing framework to evaluate the most popular CMPs and cookie opt-out tools of websites. 


Another area that has gained the attention of many researchers is web tracking and obfuscation of privacy-related information. Sanchez et al.~\cite{sanchez2018knockin} proposed an automatic tool (TrackingInspector) to detect and distinguish between known and unknown types of web tracking scripts among the Alexa top 1M websites. Mehrnezhad et al.~\cite{mehrnezhad2020cross} analyzed the privacy notices and tracking behaviours on three platforms: PC browser, mobile browser, and mobile apps and demonstrated that most web pages present the privacy consent banner inconsistently across the three platforms. They also discovered that websites collect user information before users approve on the privacy consent. Another research leveraged a crawler to investigate 2.5M clicks to seek signs of misbehaviour on various websites~\cite{sanchez2020dirty} and demonstrated that most websites hindered users from making a decision based on their intention by hiding the links' real aim or providing insufficient information. The same type of practices was also found in mobile apps by Zimmeck et al.~\cite{zimmeck2019maps}, analyzed 1 million Android apps to identify the privacy policies these apps follow and how compliant the apps are towards them.


Researchers have also developed survey templates to gather data based on users' perspectives pertaining to areas such as data deletion, email communication opt-outs and targeted advertising opt-outs. For example, results from a survey study conducted by Habib et al.~\cite{habib2019empirical} showed that the usability of these opt-outs needs improvement in several aspects, like display location and link quality. Another study by Redmiles et al.~\cite{redmiles2020comprehensive} explored the quality of 374 security and privacy advice pieces through their proposed dimensions: comprehensibility, actionability, and efficacy. Their findings revealed that though users can be aware of and understand most security advice, they needed guidance to prioritize these pieces of advice. 


\subsection{Privacy Research in the Health Sector}


Researchers have been using several techniques to evaluate the degree of privacy health websites provide. The US Department of Health and Human Services' Office of Disease Prevention and Health Promotion (ODPHP) designed a survey template to assess the privacy usability of the 100 top-ranked health-related websites~\cite{devine2016making}. This paper built a standardized quality measurement method and benchmarks to measure and capture the current status of website quality. Then based on their findings and previous research, they proposed improvement suggestions for websites. However, the template focuses on the general usability principles of a website rather than privacy-related usability. Boon-itt et al.~\cite{boon2019quality} utilized the technology acceptance model (TAM) to classify health website quality attributes and demonstrated how the quality influences the user intention to use the health website. 

Ermakova et al.~\cite{ermakova2015readability} provided evidence on the poor readability of privacy policies on health and e-commerce websites and demonstrated that privacy policies on health websites are more readable than e-commerce ones. This paper also provided an informative corpus of privacy policy texts for further research. In 2020, Ali et al.~\cite{ali2020readability} discussed the readability issues in privacy policies through literature reviews. This paper compared research results on different readability measurements and grouped them into two categories, from the reader's perspective and policy text content’s perspective. Sangari et al.~\cite{sangariidentifying} analyzed 12000 usernames and their profiles of a medical forum (Breastcancer.org) and demonstrated that patients' medical privacy was at significant risk. This paper suggested that medical forums provide privacy notices for patients before creating usernames. 

In summary, current research on privacy in health usually concerns data protection measures employed by websites or apps rather than privacy controls for users. A Few articles discussed privacy controls like privacy policy, but there is a lack of comprehensive privacy control research in the health field.

\subsection{Previous Usability Definition}
In 1994, Nielsen~\cite{nielsen1994usability} gave a general definition for usability, `a quality attribute that assesses how easy user interfaces are to use.', and described it from five dimensions: learnability, efficiency, memorability, errors, and satisfaction. In 1998, International Standard Organisation (ISO) 9241-11~\cite{international1998iso} defined the usability from three dimensions, that is effectiveness, efficiency and satisfaction. In 2010, ~\cite{kainda2010security} described usability from six dimensions, which combined ~\cite{nielsen1994usability} and~\cite{international1998iso}. In 2019, ~\cite{bt2019review} reviewed research works from 2000 to 2018 and found that it is challenging to design a unified evaluation framework to measure all the usability dimensions for e-commerce websites. In 2021, ~\cite{albesher2021evaluating} proposed a new definition of usability for all fields of study, `an evaluation of the level of quality of a user’s experience (UX)', from awareness, choice, security and redress attributes. About privacy controls, ~\cite{garlach2018m} described the usability for AdChoice icon from four dimensions: 1) time to find icon; 2) icon visibility; 3) icon functionality; and 4) adjusting preferences. Then Habib et al.~\cite{habib2020s, habib2022okay} designed usability definitions for privacy choice and cookie consent interfaces from four and seven dimensions, respectively. 

%% file: Sections/Background.tex
\section{Background}
\label{sec:background}

\subsection{Usability Definition}

Applying a single standard definition of usability across various domains is difficult since it is usually contextual. However, to analyze privacy controls, we need to collect related information that can be utilized to represent usability. One such definition in the privacy field is evaluating the quality level of a user’s experience on the four attributes notice/awareness, choice/consent, integrity/security, and enforcement/redress ~\cite{albesher2021evaluating}. The usability under specific privacy controls such as the AdChoice icon, can be evaluated under four attributes: time to find the icon, icon visibility (i.e., size, position, state, color), icon functionality, and adjusting preferences~\cite{garlach2018m}. For privacy choices, the cognitive and physical processes required to use privacy choices (i.e., opt-out and data deletion) can be utilized to represent usability~\cite{habib2020s}. This was done by designing four Human-Computer Interactions (HCI) based on finding, learning, using, and understanding a privacy policy. And for the usability of cookies, previous research have measured the usability by consent interfaces, inclusion of user needs, ability and effort, awareness, comprehension, sentiment, decision reversal, and nudging patterns~\cite{habib2022okay}.

\subsection{Privacy Controls}



A \textbf{privacy nudge} should provide both privacy information and the option for privacy choices to the users. Users should have the ability to respond for a privacy nudge by ticking check boxes, pressing a button, or toggling the selection switch. Generalized nudges can be found in various forms, for example, presentation, information, defaults, and incentives~\cite{ioannou2021privacy}. They can also be grouped into six different categories based on the task they are facilitating~\cite{schobel2020understanding}: defaults, presentation, information, feedback, error, social influence, and progress bar. Whereas, the \textbf{privacy notice} only provides the user with privacy information. There is no option for the user to take an action on the displayed notice except for removing it from their display. The \textbf{privacy policy} is a statement explaining what information is collected and how a website collects and processes the user's personal information in a relatively fixed text structure. Finally, \textbf{privacy setting} is a tool that allows users to control their data privacy according to their service demands and personal willingness.

\begin{figure}[htb!]
\centering
\subfigure{\includegraphics[width=0.3\textwidth,trim={0 -1cm -1cm 0}]{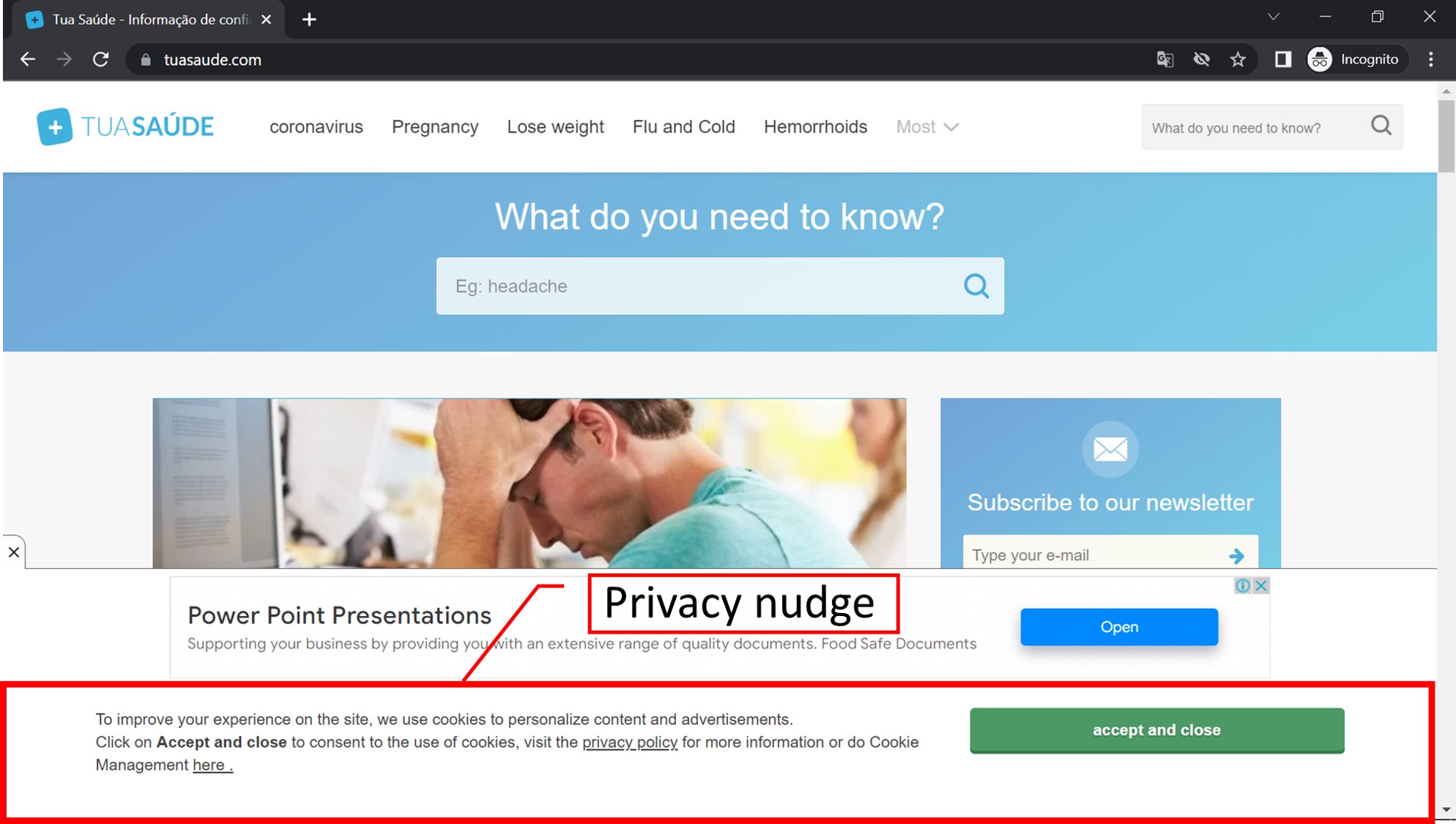}}
\subfigure{\includegraphics[width=0.3\textwidth,trim={-0.1cm -1cm -1cm 0}]{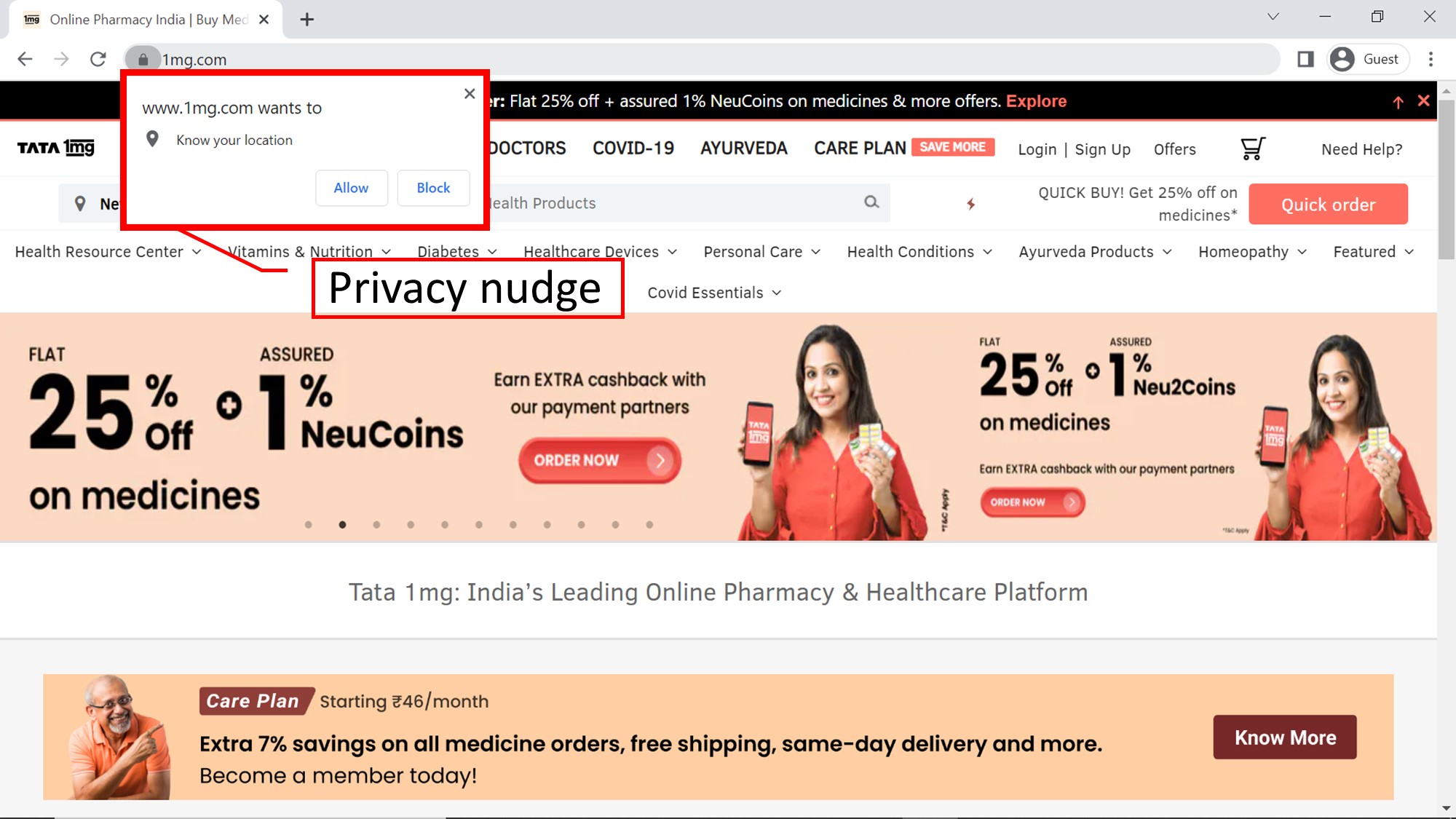}}
\subfigure{\includegraphics[width=0.3\textwidth,trim={-0.1cm -1cm -1cm 0}]{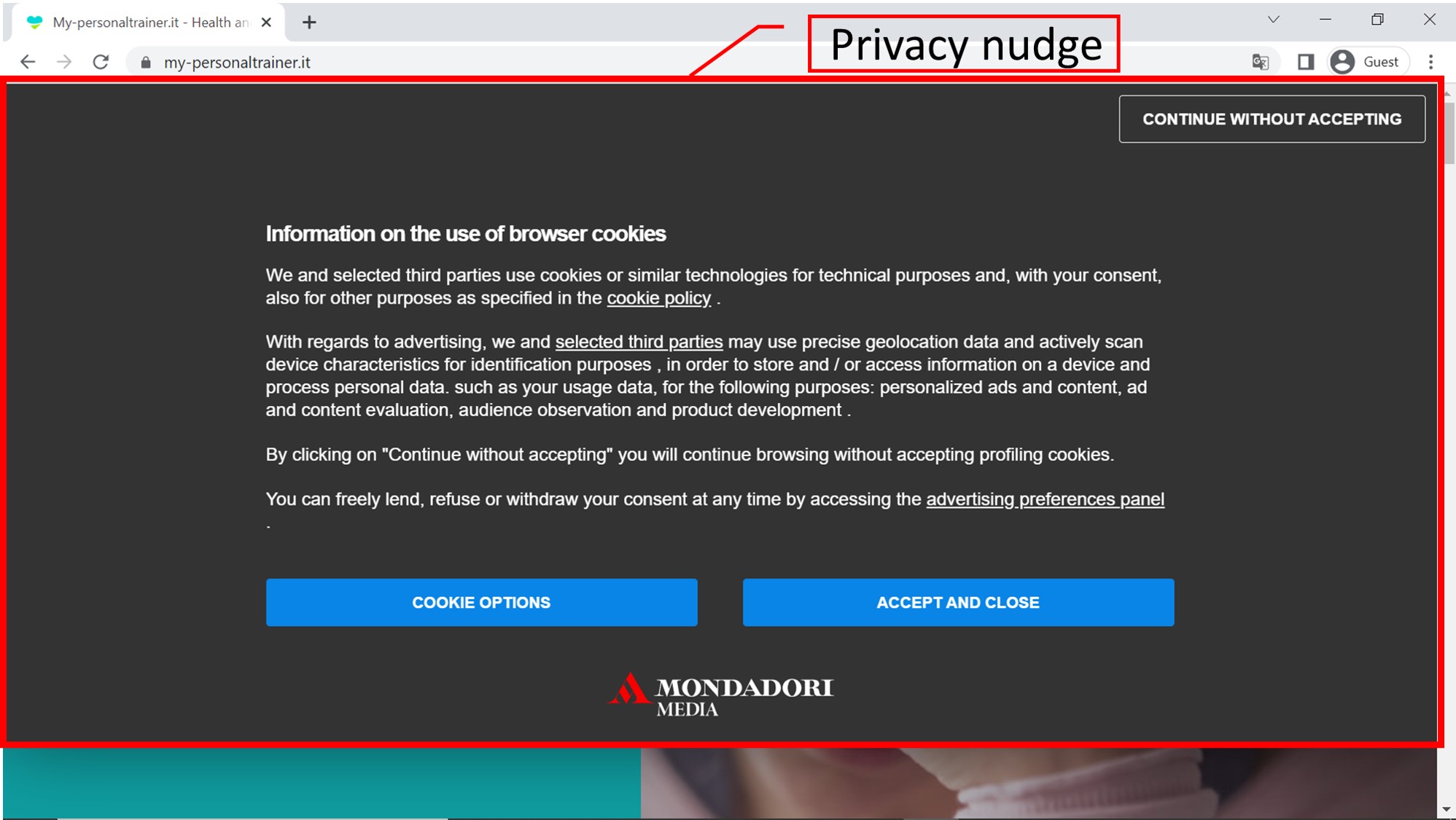}}
\subfigure{\includegraphics[width=0.3\textwidth,trim={0 -1cm -1cm 0}]{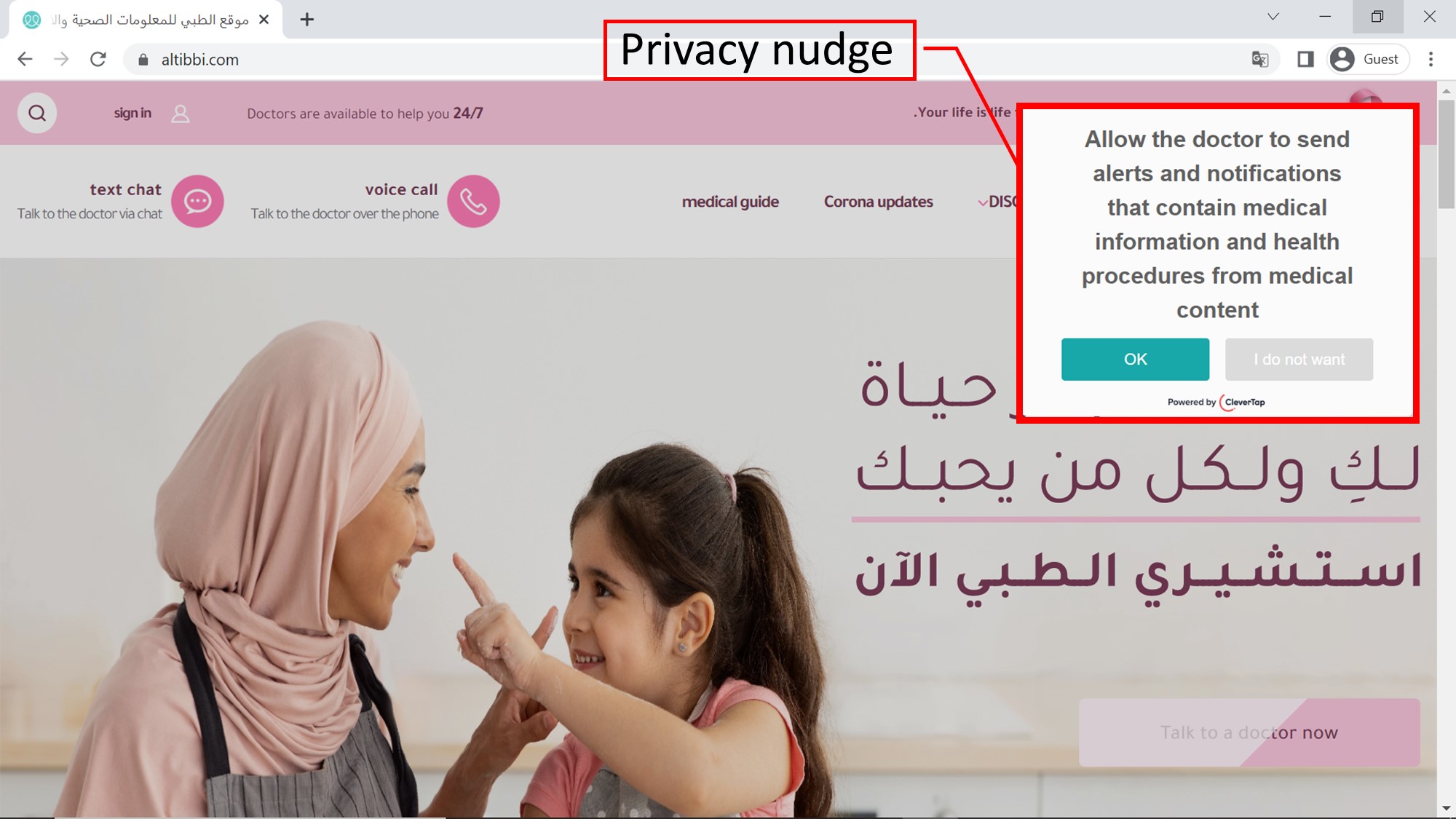}}
\subfigure{\includegraphics[width=0.3\textwidth,trim={-0.1cm -1cm -1cm 0}]{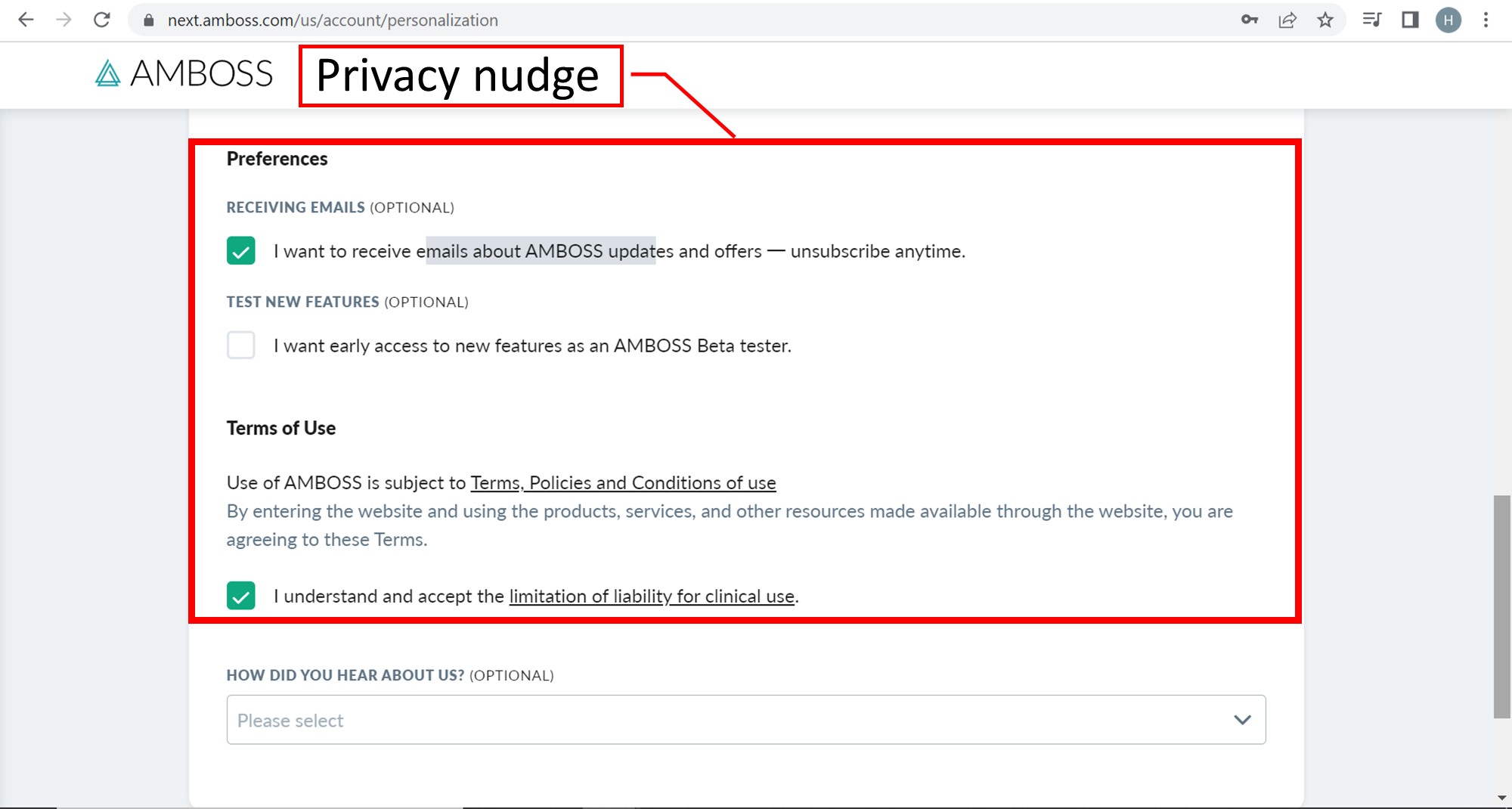}}
\subfigure{\includegraphics[width=0.3\textwidth,trim={-0.1cm -1cm -1cm 0}]{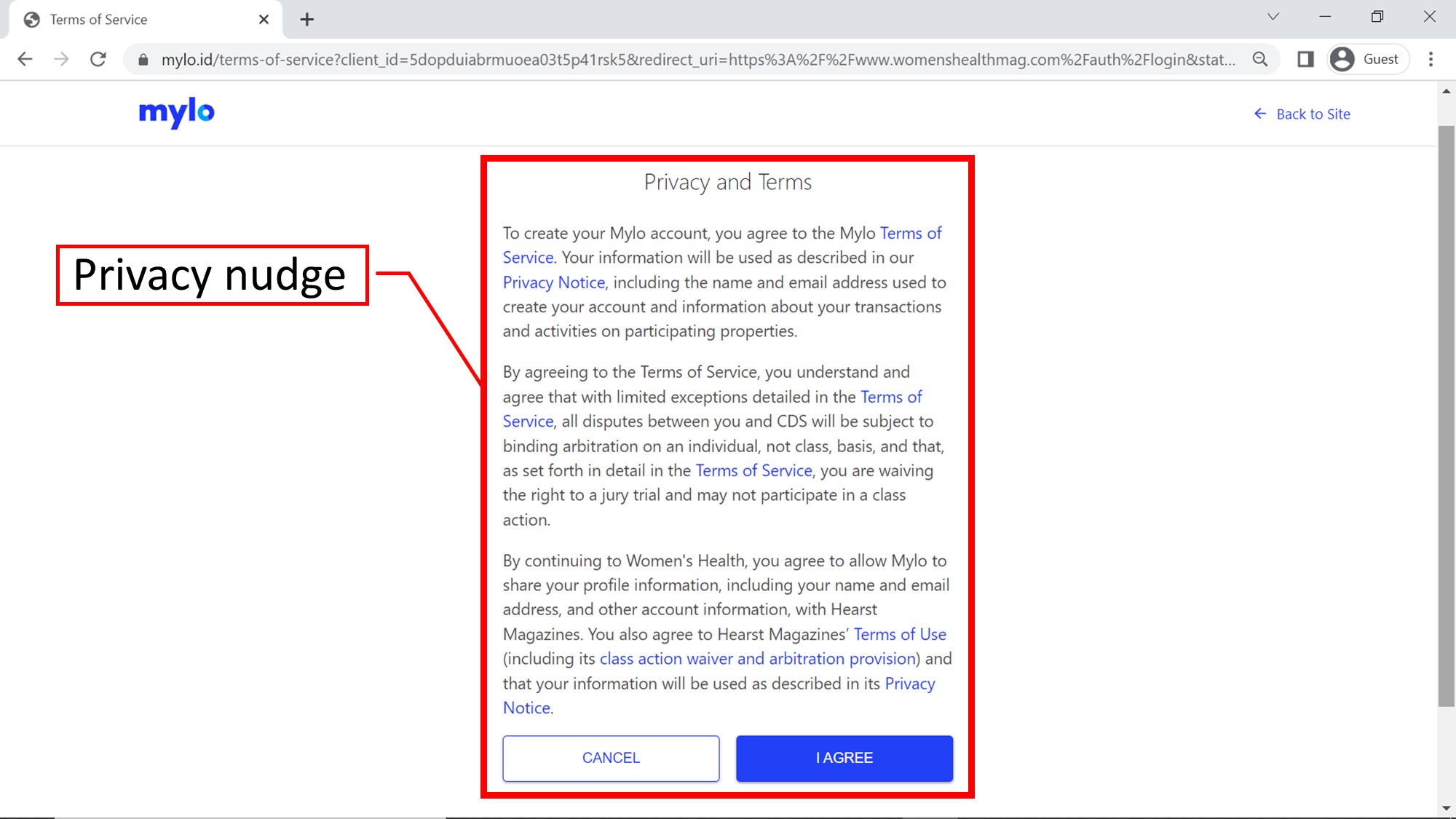}}
\caption{Privacy Nudge Examples}
\label{fig:nudge}
\end{figure}

\begin{figure}[htb!]
\centering
\subfigure{\includegraphics[width=0.3\textwidth,trim={0 -1cm -1cm 0}]{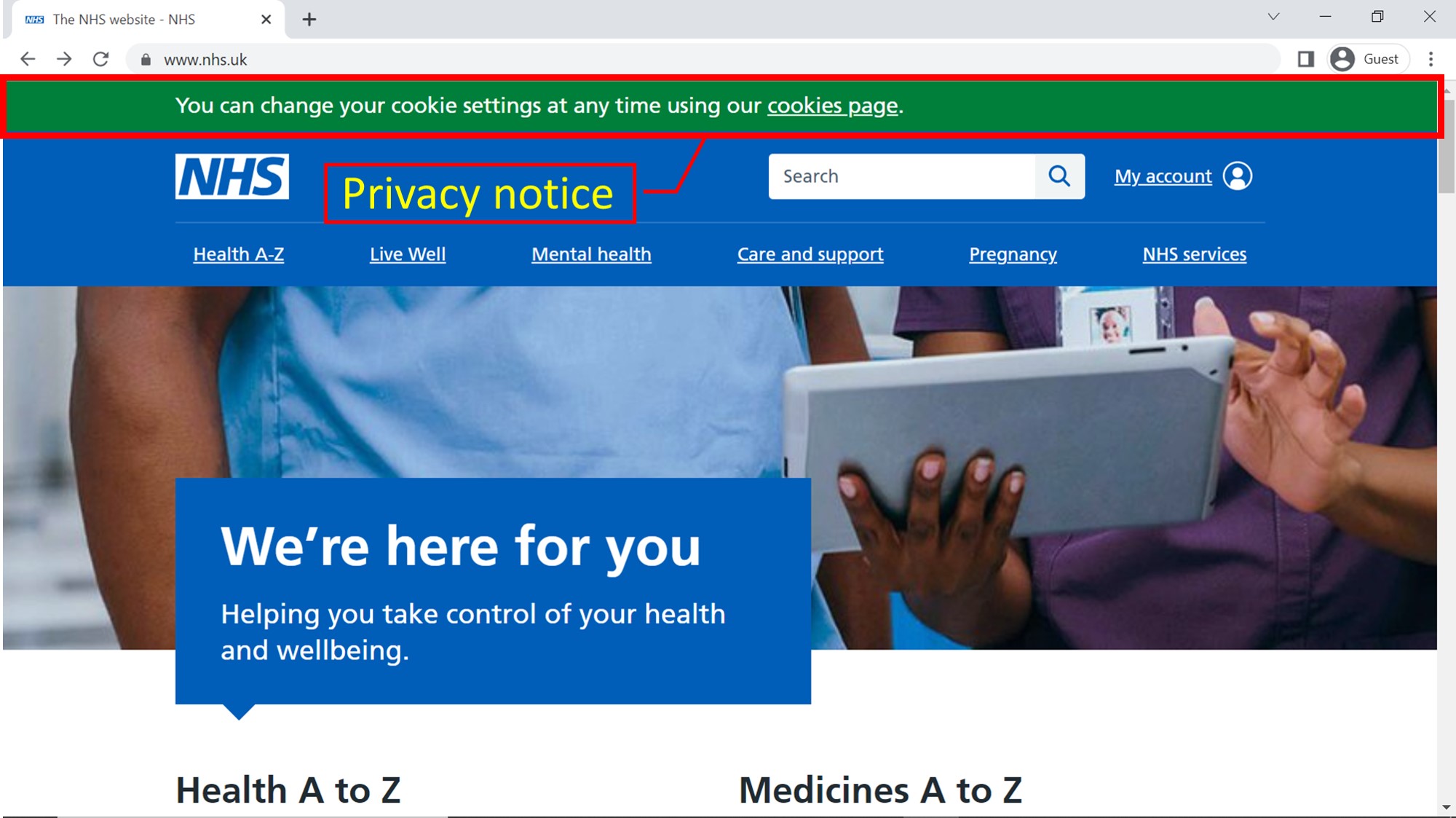}}
\subfigure{\includegraphics[width=0.3\textwidth,trim={-0.1cm -1cm -1cm 0}]{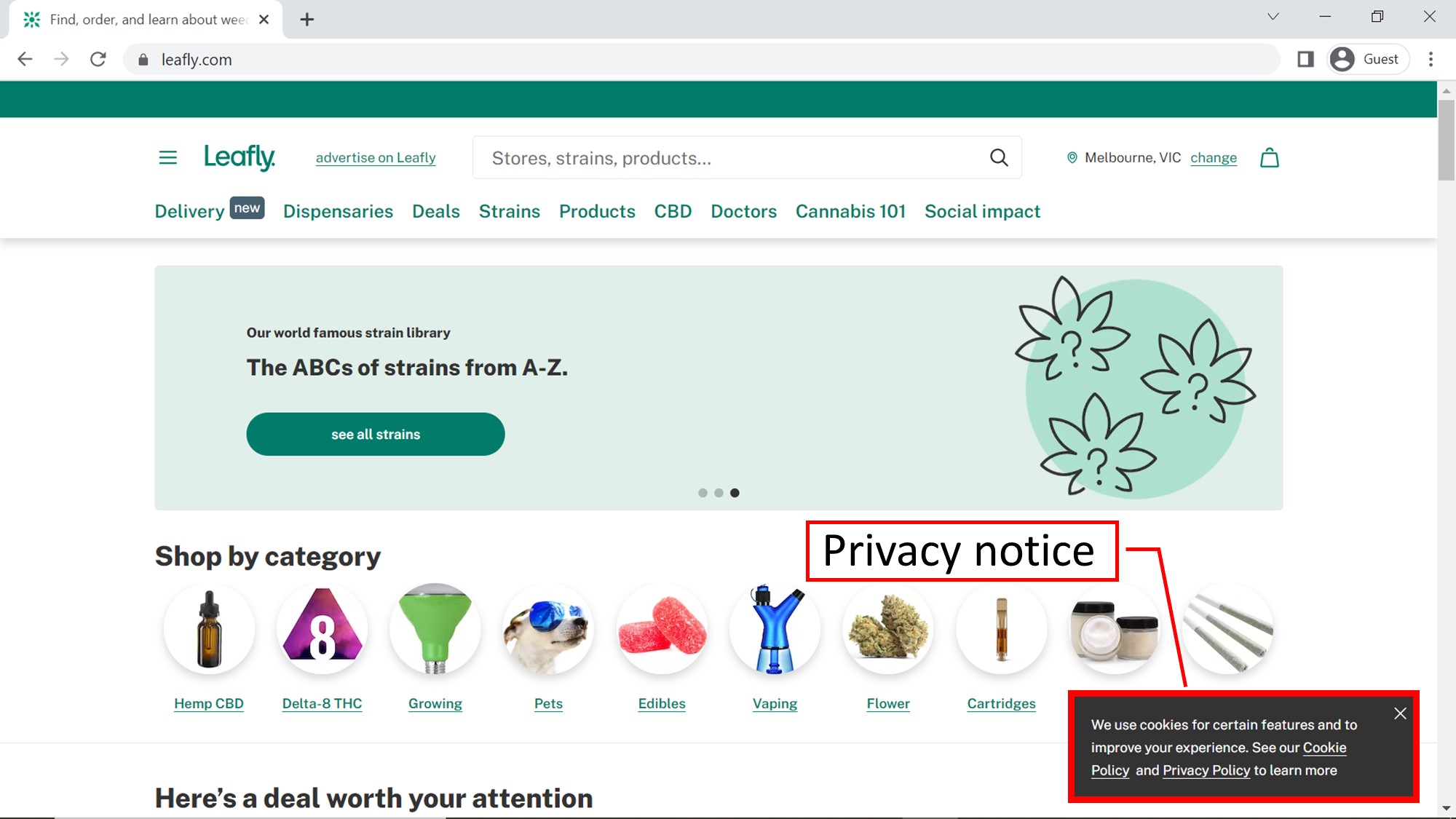}}
\subfigure{\includegraphics[width=0.3\textwidth,trim={-0.1cm -1cm -1cm 0}]{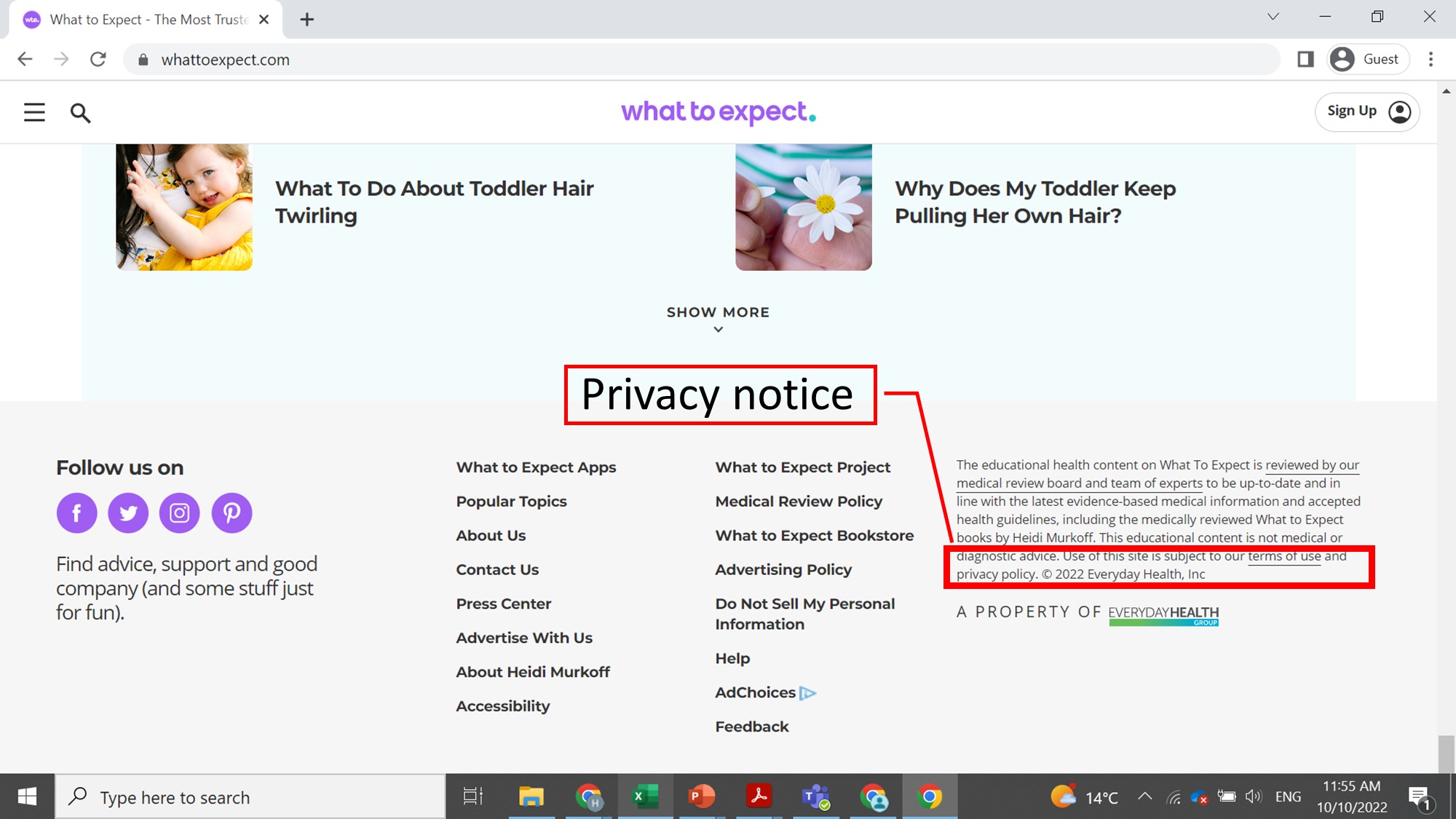}}
\caption{Privacy Notice Examples}
\label{fig:notice}
\end{figure}

Figure~\ref{fig:nudge} depicts the location of the privacy nudge in six different websites. From these six images we can clearly see that there is no fixed location for the nudge to be located and it could be present anywhere on the screen. Some of these nudges are in the form of a banner, whereas some are presented as pop-up windows. According to previous research~\cite{chatterjee2008unclicked}, pop-up windows are able to attract more user attention due to their intrusive exposure format compared to the voluntary exposure format of banners. In addition, even within banners, full screen banners are more likely to attract user attention than window/stripe banners. Another factor that plays a major role in gaining the attention of the user to a banner is the relative location of it. Banners that are located on the top of the websites obtain better visual attention compared to the bottom-right or bottom-left positions~\cite{munoz2021influence}. This behaviour is almost identical to privacy notices as well, which can be seen under Figure~\ref{fig:notice}. Hence, to evaluate the awareness of privacy nudges and notices, both the display type(pop-up/banner) and the location where it is located should be considered. 


For the privacy policy and setting, instead of the display type, the number of clicks that is required to access the feature can be utilized to measure the ease of use. This is evaluated under the efficiency attribute under the guest and log-in visit scenarios. 

The privacy policy and privacy settings should be comprehensible to the users for them to make informed decisions. Comprehension can be measured through two finer features: language and table of content. Providing multiple languages helps certain users understand the complex and professional information in the privacy policy and privacy setting with their own languages. Which can be used to estimate the quality of the website content design ~\cite{devine2016making,chiew2003webuse}. The table of content acts as a feature to represent the readability of a privacy policy because it provides a shortcut to the content that users care about the most~\cite{habib2019empirical}. In contrast, some websites purposefully degrade the comprehensibility of their privacy policy and privacy settings through dark patterns. These patterns such as scarcity, urgency, information hiding, and asymmetric choices~\cite{rinehart2022bringing} are designed to trick users by misguiding or misinforming them. However, we can assume experts in the privacy domain can identify most dark patterns and not be misled by them during the survey. The types of options that are available under privacy settings can also influence/force users to neglect their privacy rights. For example, users usually give the consent without thorough thinking when the cookie consent only provide ‘Accept’ choice. In contrast, the explicit choice guidance feature respects a user's privacy rights and tries to provide the most suitable privacy setting customization for each user.

%% file: Sections/Methodology.tex
\section{Methodology}
\label{sec:methodology}

In this section we describe the process we followed to create a template to evaluate the usability attributes Awareness, Efficiency, Comprehension, Functionality and Choice across the four different privacy controls nudge, notice, policy, and setting. Then we will talk about how we utilized this template to collect data from 100 health websites. 


\subsection{Survey Template Design}

We designed five attributes for each privacy control which represent the ease of use (\textit{awareness}, \textit{efficiency} and \textit{comprehension}) and their utilities (\textit{functionality}). In addition, we consider the influence of dark patterns on the usability of these privacy controls through \textit{choice} attribute. We use the number of clicks to access the privacy control and the type of privacy content to represent the efficiency and functionality, respectively. We utilize two finer features to represent awareness, comprehension, and choice attributes. A comprehensive breakdown of the usability features taken into consideration under each privacy control is shown in Table~\ref{table:usability}.

\begin{table}[htb]
\caption{Usability Attributes for Four Privacy Controls in Three Visit Scenarios}
\label{table:usability}
\footnotesize
\centering
\tabcolsep=0.11cm
\begin{tabular}{lllllllllll}
\toprule
\multicolumn{2}{c}{\multirow{2}{*}{\textbf{Usability Attributes}}}	& \multicolumn{4}{c}{\textbf{Guest Visit}}	& \multicolumn{4}{c}{		\textbf{Registering}}		&	\textbf{Log-in Visit}\\
\cline{3-11}
& &\textbf{nudge}	& \textbf{notice}	&		\textbf{policy}		&	\textbf{setting} &\textbf{nudge}	&			\textbf{notice}	&		\textbf{policy}		&	\textbf{setting} &	\textbf{account setting}\\
\midrule
\multirow{2}{*}{Awareness} 	&	Location	&	\checkmark	&	\checkmark	&	\checkmark	&	\checkmark	&	\checkmark	&	\checkmark	&	\checkmark	&	\checkmark	&	\checkmark	\\
		&	Display type	&	\checkmark	&	\checkmark	&		&		&	\checkmark	&	\checkmark	&		&	\checkmark	&		\\
	Efficiency	&	Number of clicks	&		&		&	\checkmark	&	\checkmark	&		&		&		&		&	\checkmark	\\
\multirow{2}{*}{Comprehension}	&	Language	&		&		&	\checkmark	&	\checkmark	&		&		&		&	\checkmark	&	\checkmark	\\
		&	Table of content	&		&		&	\checkmark	&		&		&		&		&		&		\\
	Functionality	&	Type of privacy content	&	\checkmark	&	\checkmark	&		&	\checkmark	&	\checkmark	&	\checkmark	&		&	\checkmark	&	\checkmark	\\
\multirow{2}{*}{Choice}&	Types of options	&	\checkmark	&		&	\checkmark	&		&	\checkmark	&		&		&		&		\\
		&	Explicit choice guidance	&	\checkmark	&		&	\checkmark	&		&	\checkmark	&		&		&		&		\\
\bottomrule
\end{tabular}
\end{table}

Based on these usability attributes, we designed pertinent questions for users in the survey. There are three website visit scenarios we are interested in: guest visit, registering, and log-in visit. The survey contains three separate modules corresponding to each of the visit scenarios. Under each module, there are four sections pertaining to each privacy policy we aim to evaluate. Finally, several subsections contain specific questions for each privacy control under each section. Following the natural flow of a user surfing the web page, we set the guest visit scenario as the first module and registering (i.e., creating an account) and log-in visit scenarios as the second and last module, respectively and the privacy controls are ordered as privacy nudge, privacy notice, privacy policy, and privacy setting. 



In the first module, the first question under each privacy control is whether the user can access the relevant privacy control while visiting the web page. These questions take the simple format of a `Yes' or `No' answer. This question serves as a statistical measure to represent the availability of each privacy control. If answered `Yes', the user is prompted with the questions under each subsection of the privacy control; else, the questions are skipped and the survey moves on to the next section. Each subsection under the privacy controls consists of open questions based on `What', `Where', `How many', and `Which'. However, it is impossible to assume all available answer options before we complete a single round of data collection. Therefore, we used the Excel data validation tool to create a drop-down list to record the seen options and avoid inconsistent options for the same answer. This enables us to maintain the scalability of the template while continuously updating the available options.


Since the privacy nudge and the privacy notice follow a similar format, they share the same questions under each subsection. They are, 1) display location (`Where'), e.g., `Top', `Bottom' and `Full screen'; 2) display/trigger type (`What'), e.g., `Pop up Window', `Banner', and `Check Box Content'; and 3) privacy content/function type (`What'), e.g., `Cookie', `Advertisement', `Profile/Personally Identifiable Information (PII). Based on the answers for display location and trigger type, we summarize the frequently-used types of display and analyze their shares in the top 100 health websites. We can have a general catalog of the privacy information from the content type answers and analyze their functionality. However, there are two more questions for privacy nudge. One is about the response action type, and another is about whether there are detailed explanations for those actions. Both of these questions are used to estimate the efficacy of the privacy control. 


Under the privacy policy section, we designed nine questions. The first four questions are used for estimating the ease of use, the subsequent three are used to estimate the utility, and the last two are leveraged for regulation influence and region distribution analysis. The first three questions are: 1) where the privacy policy button is displayed; 2) how many clicks are required to visit the privacy policy; 3) whether it is available in various languages. Since the privacy policy is a legal document containing many domain-specific terms and concepts, not everyone will be able to grasp the meaning at once. Therefore, it should help the user understand the content in their familiar language. Question 4), whether it contains contents to guide reading, is used to evaluate the readability of the privacy policy to some extent. Questions 5 to 7 focus on whether the privacy policy contains: 1) privacy setting options; 2) privacy setting links; 3) clear privacy setting guidance. These three questions are designed to see whether the privacy policy provides customized rights to users. The final two questions are about the regulations this privacy policy follows and the country this website belongs to. The options  for open questions under these two sections can be found in Table~\ref{table:option1}.


For the privacy setting section, there are four questions altogether. However, the first three questions are similar to the ones on the privacy policy. The final question is based on the coverage of the privacy settings provided. The question is functionally identical to the third question of the privacy nudge and determines how many aspects the privacy setting covers. 
Finally, if the website allows users to register, the questions related to this scenario are available under the second module of the survey template. The question in these modules are nearly the same as in the first module, except for the third-level questions about the privacy policy. The content of the privacy policy does not change under each visit scenario. Hence, we only include the first question about the privacy policy button/link display location that may differ in the registering scenario. Once the user successfully registers, we move on to the final module of the survey. Since most websites only allow to change privacy settings/preferences tied to a user account, in this section we concentrate primarily on the privacy setting aspects of the web page. The options for open questions under the privacy policy and privacy setting can be found in table~\ref{table:option2}



 
\begin{table}[htb]
\caption{Options for Open Questions About Privacy Nudge/Notice}
\label{table:option1}
\footnotesize
\centering
\begin{tabular}{lllll}
\toprule
\textbf{Privacy nudge/notice location}	&			\textbf{Privacy nudge/notice triggers}	&		\textbf{Privacy nudge actions}		&	\textbf{Privacy nudge/notice  types}\\
\midrule
top	&			pop up window	&		accept	&		cookie\\
middle&				new page&			decline	&		advertisement\\
bottom&				banner&			manage&			privacy policy\\
left&				check box content& 		&				location\\
right&				plain text& 			&			email/notification\\
full screen	&			pull down menu&		&				profile/PII\\
nearby sign up/log in/subscribe&				button/link&		&		history\\
email confirming pages&		&				&				security guarantee\\ 
\bottomrule
\end{tabular}
\end{table}

\begin{table}[htb]
\caption{Options for Open Questions About Privacy Policy and Privacy Setting}
\label{table:option2}
\footnotesize
\centering
\begin{tabular}{llll}
\toprule
\textbf{Privacy policy button location}	&	\textbf{Whether the privacy policy contains the privacy setting link?} & \textbf{Privacy setting types}\\
\midrule
footer&		Yes, third party link&						cookie\\
full screen&		Yes, customized link.&						advertisement\\
link in the privacy notice/nudge&		Yes, both kinds of link.	&					location\\
&		No.&						email/notification\\
& &								profile/PII\\
& &								history\\
& &								activity \\
\bottomrule
\end{tabular}
\end{table}

\subsection{Data Collection}

During the template creation stage, we only employed one researcher to explore the websites to keep the recorded data consistent. Once the first iteration of the template was completed, the researcher browsed ten websites and recorded the data collection process according to the template. Based on these findings, we finalized the three modules of the template with their relevant sections and subsections. This template was used for the researcher's data collection of the 100 websites. We then checked the template for questions that have the same answer for all websites and removed these questions since they do not provide any new information regarding the websites. The options for the open-ended question were also finalized during this stage. Finally, two additional researchers were given the template and asked to gather information independently, giving us three independently gathered data samples. 


During the data collection process, we followed rigorous instructions to ensure outside factors did not interfere with the data gathered. The researchers were free to pick any browser for data collection since it is more realistic than every user using one browser. However, the browser should be in incognito mode with previous cookies cleared to avoid any interference from previously collected cookies. An example of such behaviour is displaying the privacy nudge/notice. If we have visited a website before with the same browser with saved cookies, they may not appear again. For the account creation process, the researchers were provided with a set of mock personal data and an email address to be used. 

On average, each researcher spent about 20 minutes per website with some sites requiring around 50 minutes. The time taken was most dependent on whether the website contained a registering function and the complexity of its privacy policy.  For example, suppose there is no sign-up option. In that case, the researcher only needs to explore the questions in the first module (i.e., guest visit scenario), which usually takes 10 minutes. On the other hand, if a website has registering option, one of two outcomes can occur. First, a lack of specific information, such as a local phone number or residential healthcare card, may result in a failed registration. Since a successful registration leads to the log-in scenario, this would result in a longer data collection duration. Finally, to answer the questions about the privacy policy, the researcher needs to skim through the whole document. While some privacy policies only have several pages, some contain over thirty pages. For a website with three visit scenarios and a complex privacy policy, a researcher may take about 50 minutes for the complete process.

\subsection{Usability Analysis Methods}
We summarize the usability analysis methods into two main categories. One is based on the real user's perspective, which needs employing many potential users to complete surveys for several websites. Most questions of this kind of survey are subjective, like 'Whether you are easy to find the privacy policy button?' or 'How you satisfied with this privacy setting choices?'. The differences in their subjective feelings are rated based on the Likert scale method (e.g., five-point or seven-point scale). According to the users' rating scores, previous research proposed different models to fit their scores from multiple feature dimensions. These models are usually based on the concept of the Structure Equation Model (SEM). The number of survey participants is the key factor in ensuring the robustness of a model. This usability analysis aims to build a measurement standard to evaluate each website.

On the other hand, another kind of usability analysis is based on the expert's perspective, which only needs one or several experts to complete surveys on numerous websites. Most questions in this kind of survey are objective (see Table~\ref{tab:template}). Thus experts can perform more professionally and consistently. In this situation, previous research usually focused on the statistical overview of the usability for certain privacy controls on all targeted websites. They analyzed each usability attribute separately with traditional statistical methods and gained insights from these results. 

In this paper, we will use expert perspective-based usability analysis approaches and conduct the usability analysis for privacy nudge, privacy notice, privacy policy, and privacy setting on awareness, efficiency, comprehension, choice, and functionality attributes.

%% file: Sections/Results.tex
\section{Results}
\label{sec:results}

\subsection{Overview of Targeted Websites}

Our research centers on the usability analysis of four privacy controls on the top 100 health websites with the highest traffic levels. Our initial selection of the 100 health websites was subject to domain restrictions, which included three inaccessible websites and two adult websites. To resolve these constraints, we incorporated an additional set of five websites sourced exclusively from Australia into our sample.

\subsubsection{Regional Distribution}

First, we examine the regional distribution of the 100 targeted health websites and collect statistical data of website counts in each country in descending order (see Table ~\ref{table:website_rank}).

Our study revealed that the majority of the websites in our sample were hosted in the United States of America (46\%), followed by Russia (6\%), Indonesia (6\%), the United Kingdom (6\%), Australia (6\%), Brazil (4\%), India (3\%), and Canada (3\%). In the targeted set of 100 websites, the remaining thirteen countries only host one or two websites that are included.

\begin{table}[h]
\caption{Website Regional Distribution}
\label{table:website_rank}
\footnotesize
\centering
\begin{tabular}{  lll }
\toprule
\textbf{Rank}&\textbf{Website Count}&\textbf{Country Name} \\
\midrule

1&	46&	USA\\
2&	6&	Russia, Indonesia,UK and Australia.\\
3&	4&	Brazil\\
4&	3&	India and Canada\\
5&	2&	France, China, Germany, Mexico, Poland and Japan\\
6&	1&	Vietnam, Ukraine, Portugal, Italia, UAE, Jordan, Netherlands and Global\\
\bottomrule
\end{tabular}
\end{table}

Furthermore, the identified websites have been classified into six groups based on their respective geographic regions: North America, Europe,
Asia-Pacific, South America, Middle East, and Global. The global website means it belongs to an international organization, not a specific country or region, e.g., the world health organization. The geographical distribution of the 100 websites is shown in Figure 3. We can obverse that North America, Europe, and
Asia-Pacific constitutes the majority of targeted websites (93\%). Therefore, we conduct research on the websites originating from the top three regions exclusively in the subsequent region-related
analysis.

\begin{figure}[h]
\begin{centering}
\label{regionaldis}
\includegraphics[width=0.6\textwidth]{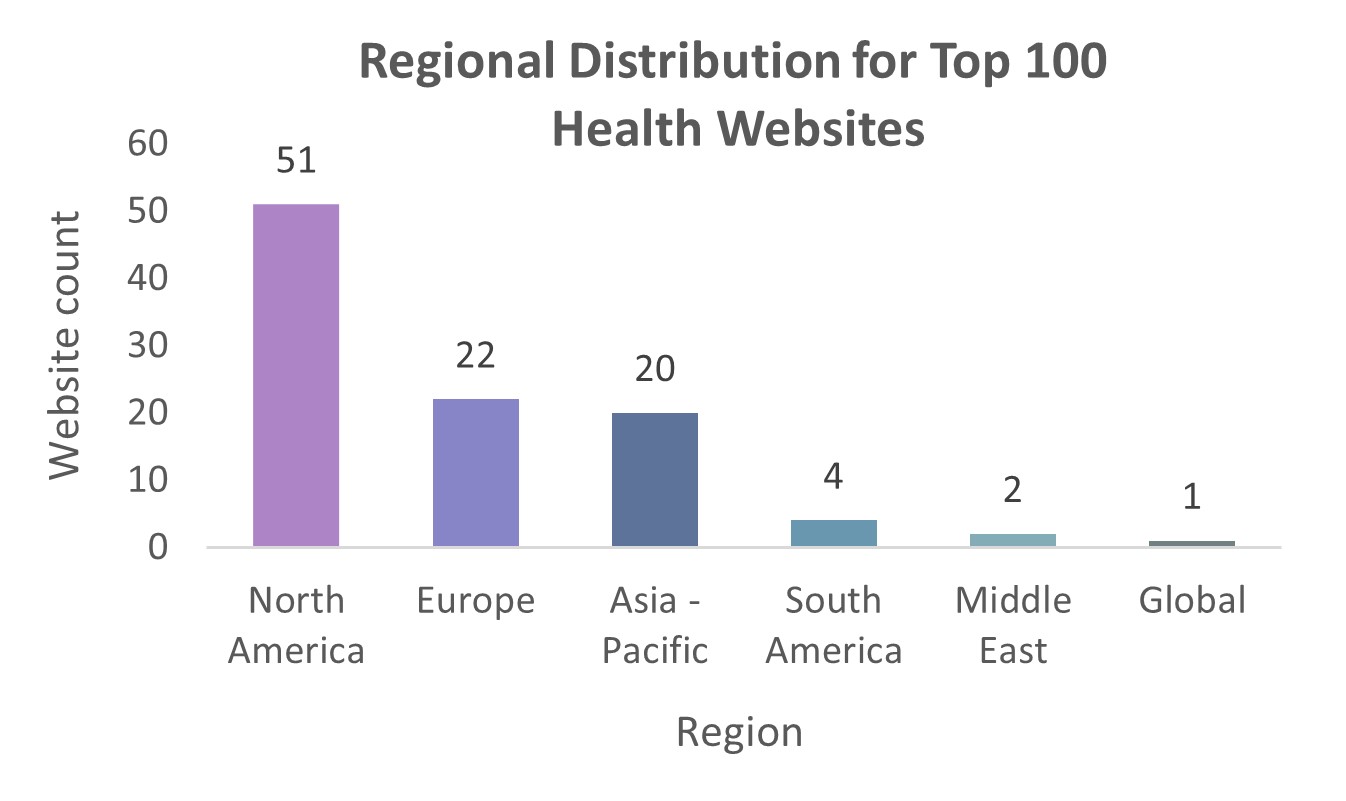}
  \caption{Geographical Distribution for Top 100 Health Websites.}
  \label{region}
\end{centering}
\end{figure}

\subsubsection{Regulation Compliance}

Following the enforcement of regulations and protocols regarding user data privacy protection, exemplified by GDPR and CCPA, it has become imperative for website designers to abide by these regulations and respect users' privacy. General Data Protect (GDPR) mainly regulates websites in Europe, while California Consumer Privacy Act (CCPA) primarily applies to California and other states of the USA. While exploring the
privacy-related regulations followed by the top 100 health websites, we notice that the degree of regulatory compliance varies across different geographical regions. North America (including the USA, Mexico, and Canada) and Europe (including Russia and Ukranine) place more emphasis on privacy regulatory compliance. 

\begin{table}[h]
\caption{The Count of Websites Complying with Each Regulation}
\label{table:regulation}
\footnotesize
\centering
\scalebox{0.93}{
\begin{tabular}{  llll }
\toprule
\textbf{Regulation Abbreviation}&\textbf{Regulation Full Name}&\textbf{Country} &\textbf{Count} \\
\midrule
CCPA 	&	California Consumer Privacy Act	&	USA	&	25	\\
GDPR	&	General Data Protection Regulation	&	Europe	&	18	\\
HIPAA	&	Health Insurance Portability and Accountability Act Notice of Privacy Practices 	&	USA	&	11	\\
Act	&	Privacy Act 1988  	&	Australia	&	6	\\
Law No. 152-FZ	&	The Federal law of July 27 2006 No. 152-FZ	&	Russia	&	5	\\
LGPD	&	Lei Geral de Proteção de Dados (General Data Protection Law)	&	Brazil	&	4	\\
PA	&	Privacy Act of 1974 USA	&	USA	&	3	\\
CCC	&	Califorlia Civil Code	&	USA	&	2	\\
APEC CBPR	&	Asia-Pacific Economic Cooperation Cross Border Privacy Rules Program Requirements.	&	USA	&	2	\\
COPPA	&	Children’s Online Privacy Protection Act	&	USA	&	2	\\
LFPDPPP	&	Ley Federal de Protección de Datos Personales en Posesión de los Particulares (Federal Law &	Mexico	&	2\\
& on Protection of Personal Data Held by Private Parties)	& &	\\
NL	&	Nevada Law	&	USA	&	1	\\
FOIA	&	The Freedom of Information Act and Privacy Act	&	USA	&	1	\\
PIPEDA	&	Personal Information Protection and Electronic Documents Act	&	Canada	&	1	\\
SPI	&	Regulation 4 of the Information Technology (Reasonable Security Practices and Procedures&	India&	1\\
&and Sensitive Personal Information) Rules, 2011 (the “SPI Rules”)	& &		\\
APP	&	Australian Privacy Principles	&	Australia	&	1	\\
PCR	&	The Law of Ukraine "On Protection of Consumer Rights"	&	Ukraine	&	1	\\
APPI	&	Act on the Protection of Personal Information 	&	Japan	&	1	\\

\bottomrule
\end{tabular}
}
\end{table}

From Table ~\ref{table:regulation}, it can be observed that the predominant (top two) regulations that are commonly employed are the CCPA and GDPR from the North American and European regions respectively.
In addition, there are ten regulations employed by targeted websites originating from North America, three from Europe, and four from the Asia-Pacific region.
As illustrated in Figure 4, the regulatory compliance ratios in North America and Europe exceeds
80.0\%, whereas in the Asia-Pacific region, the ratio is only half of that observed in the former two regions. Taking a close look at the Asian-Pacific region, there are 20 websites from six countries and one area (China TaiWan). Nevertheless, amongst them, only 8 websites (6 from Australia, 1 from India, and 1 from Japan) comply with privacy regulations. In Section 5.2, when we analyze the usability of the privacy controls, we will attempt to further analyze the relationship between privacy control usability and the website regulatory compliance ratios of different regions.

\begin{figure}[htb]
\begin{centering}
\includegraphics[width=0.5\textwidth]{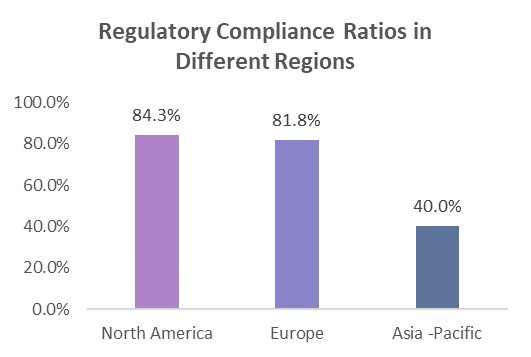}
  \caption{Regulatory Compliance Ratios in Different Regions}
  \label{regulation}
\end{centering}
\end{figure}

\subsection{Availability of Privacy Controls}

Before specific privacy control analysis, we summarize the overall availability of privacy nudge, privacy notice, privacy policy, and privacy setting in three visit scenarios (log-in visit, guest visit, and registering) in Figure 5. We conduct an analysis of all 100 websites to assess their performance in the guest visit scenario. Our findings indicate that 42 of these websites were capable of complete account creation for users, hence only these 42 websites are explored in registering and log-in visit scenarios.

The availability of each privacy control varies depending on the visit scenarios, and the privacy nudge, privacy notice, and privacy setting are not available for most websites. 
Most websites (97.6\%, 99.0\%, and 81.0\%) contain privacy policies in three visit scenarios. It is noteworthy that solely the Russian website 'sportkp.ru' does not present a hyperlink to its privacy policy on the homepage during the guest visit scenario. Nevertheless, it provides a privacy policy link during the registering scenario. That means all targeted websites offer a privacy policy that is accessible to users. Typically, the privacy policy is made available on the website homepages, and in some cases, during the registration process.

On the other hand, the availability of privacy nudges, privacy notices, and privacy setting is considerably lower. In the guest visit scenario,  34.0\% websites have privacy setting buttons, whereas only 29.0\% and 13.0\% websites apply the privacy nudge and privacy notice respectively. In the registering scenario, the availability is significantly higher as compared to the guest visit scenario, with websites having privacy nudges and privacy notices constituting 61.9\% and 64.3\%. In the log-in visit scenario, the websites with privacy nudges and notices are fairly rare, with merely 0.0\% and 4.8\% presented respectively. The availability of the privacy setting (52.4\%) in the log-in visit scenario is significantly higher than in the guest visit and registering scenarios. In summary, the privacy policy is the privacy control of the highest availability for websites in three visit scenarios. In addition, the websites tend to apply the privacy nudge and notice in the registering scenario rather than in the guest visit scenario. The privacy setting is utilized most frequently in the log-in visit scenario.

\begin{figure}[h]
\begin{centering}
\includegraphics[width=0.8\textwidth]{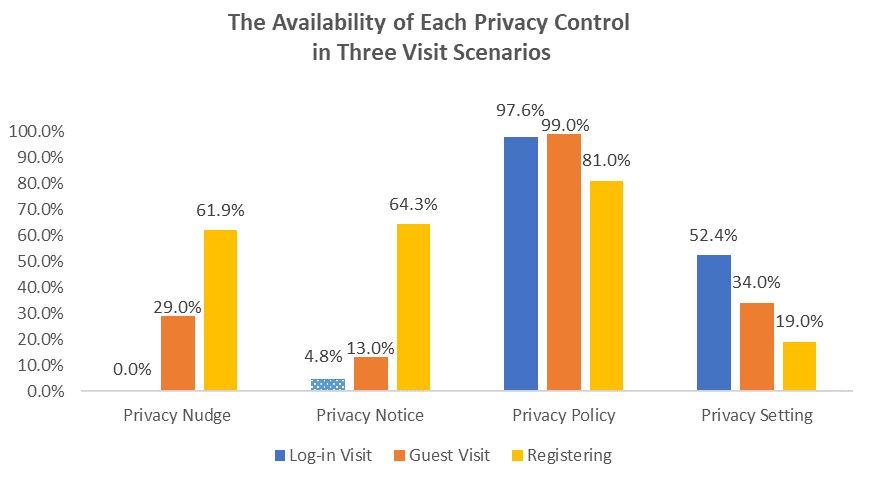}
  \caption{The Availability of Each Privacy Control
 in Three Visit Scenarios}
  \label{availability_scenario}
\end{centering}
\end{figure}

\subsubsection{Based on Regional Distribution}

We provide the availability of each privacy controls in different regions in Table~\ref{table:availability} and find the regional imbalance.
We can see that the availability of the privacy nudge, notice, and setting in any visit scenario (except the privacy nudge in the guest visit scenario) in the Asia-Pacific region are significantly lower than in the other two regions. Especially, websites hosted in the Asia-Pacific region notably lack both privacy notices (in the context of guest visits) and privacy settings (in both guest visit and registering scenarios). This finding implies not all regulations will improve the availability of privacy controls, as some regulations may lack content or clear guidance in these areas. That means the rate of regulatory compliance in a region is not necessarily directly and positively correlated with the availability of privacy controls. However, we can conclude that a low regulatory compliance ratio does result in the low availability of privacy controls from the findings in Table~\ref{table:availability}. 

\begin{table}[h]
\caption{The Availability of Each Privacy Control in Different Regions}
\label{table:availability}
\footnotesize
\centering
\begin{tabular}{  lllll }
\toprule
\textbf{Guest Visit Scenario}&\textbf{Nudge}&\textbf{Notice}&\textbf{Policy}&\textbf{Setting} \\
\midrule
The number of websites applicable & 100 &100 &100 & 100\\
The number of total websites with the privacy control &29 &13 &99 & 34 \\
The number of North America websites with the privacy control (total 51) &9 &7 &51 & 19 \\
The number of Europe websites with the privacy control (total 22) &14 &5 &21 & 14 \\
The number of Asia-Pacific websites with the privacy control (total 20) &4 &0 &20 & 0 \\
The number of other region websites with the privacy control (total 7) &2 &1 &7 & 1 \\
\midrule
\textbf{Registering Scenario}&\textbf{Nudge}&\textbf{Notice}&\textbf{Policy}&\textbf{Setting} \\
\midrule
The number of websites applicable & 42 &42 &42 & 42\\
The number of total websites with the privacy control &26 &27 &34 & 8 \\
The number of North America websites with the privacy control (total 18) &12 &14 &15 & 3 \\
The number of Europe websites with the privacy control (total 12) &9 &8 &10 & 5 \\
The number of Asia-Pacific websites with the privacy control (total 8) &4 &2 &5 & 0 \\
The number of other region websites with the privacy control (total 4) &1 &2 &4 & 0 \\
\midrule
\textbf{Log-in Scenario}&\textbf{Nudge}&\textbf{Notice}&\textbf{Policy}&\textbf{ Account Setting} \\
\midrule
The number of websites applicable & 42 &42 &42 & 42\\
The number of total websites with the privacy control &0 &2 &41 & 22 \\
The number of North America websites with the privacy control (total 18) &0 &0 &18 & 11 \\
The number of Europe websites with the privacy control (total 12) &0 &1 &11 & 7 \\
The number of Asia-Pacific websites with the privacy control (total 8) &0 &0 &8 & 2 \\
The number of other region websites with the privacy control (total 4) &0 &1 &4 & 2 \\
\bottomrule
\end{tabular}
\end{table}

\subsubsection{Based on Regulation Compliance}

While not all regulations are effective in promoting privacy controls, both GDPR and CCPA contribute to some privacy controls significantly. Even though 73 of the 100 websites comply with certain regulations, the availability of each privacy control is not consistent with the general regulatory compliance ratio (see Table~\ref{table:availability_regulation}). This may be attributed to the fact that some regulations do not require certain privacy controls. For example, our finding suggests that all targeted websites complying with HIPAA and Act do not provide either the privacy nudge and notice in the guest visit scenario, or the privacy setting in the registering and log-in visit scenario. Only 10.0\% of the above-mentioned websites provide the privacy setting in the guest visit scenario, and 20.0\% offer the privacy nudge and notice in the registering scenario.
Nonetheless, if we narrow our attention to websites that adhere to GDPR or CCPA, the data reveals that the rates of privacy setting availability are considerably greater, exceeding those of other regulations by over 10.0\% in all three visit scenarios. In addition, websites complying with GDPR have better availability of the privacy nudge in three visit scenarios while websites complying with CCPA contribute to the privacy notice in the registering scenario. 

\begin{table}[h]
\caption{Availability Comparison of Each Privacy Control for All Websites and Websites Complying With Regulations}
\label{table:availability_regulation}
\footnotesize
\centering
\begin{tabular}{  lllll }
\toprule
\multirow{2}{*}{\textbf{Visit Scenarios}} &\textbf{Nudge}&\textbf{Notice}&\textbf{Policy}&\textbf{Setting}\\
\cline{2-5}
&\multicolumn{4}{c}{All Websites/Websites Complying With Regulations} \\
\midrule
Log-in Visit	&	0.0\%	/	0.0\%	&	4.8\%	/	100.0\%	&	97.6\%	/	100.0\%	&	52.4\%	/	51.5\% \\
Guest Visit	&	29.0\%	/	27.4\%	&	13.0\%	/	15.1\%	&	99.0\%	/	100.0\%	&	34.0\%	/	41.1\% \\
Registering	&	61.9\%	/	57.6\%	&	64.3\%	/	63.6\%	&	81.0\%	/	81.8\%	&	19.0\%	/	21.2\% \\
\midrule
&\multicolumn{4}{c}{All Websites/Websites Complying With GDPR/Websites Complying With CCPA} \\
\midrule
Log-in Visit	&	0.0\%	/	0.0\%/ 0.0\%	&	4.8\%	/	50.0\%/0.0\%	&	97.6\%	/	100.0\%/100.0\%	&	52.4\%	/	\textbf{62.5\%/66.7\% }\\
Guest Visit	&	29.0\%	/	\textbf{57.1\%}/17.6\%	&	13.0\%	/	7.1\%/11.8\%	&	99.0\%	/	100.0\%/	100.0\%	&	34.0\%	/	\textbf{78.6\%/52.9\%} \\
Registering	&	61.9\%	/	\textbf{75.0\%}/\textbf{77.8\%}	&	64.3\%	/	62.5\%/\textbf{77.8\%}	&	81.0\%	/	75.0\%/88.9\%	&	19.0\%	/	\textbf{50.0\% /33.3\%}\\
\bottomrule
\end{tabular}
\end{table}

\subsection{Usability of Privacy Controls}

In this section, an usability analysis is performed according to usability attributes spanning all four privacy controls in Table~\ref{table:usability}. The following sub-sections conclude the study of the usability attributes corresponding to each privacy control in different scenarios.

\subsubsection{Privacy Nudge}

\label{nudge_analysis}
\begin{figure}[h]
\begin{centering}
\includegraphics[width=0.8\textwidth]{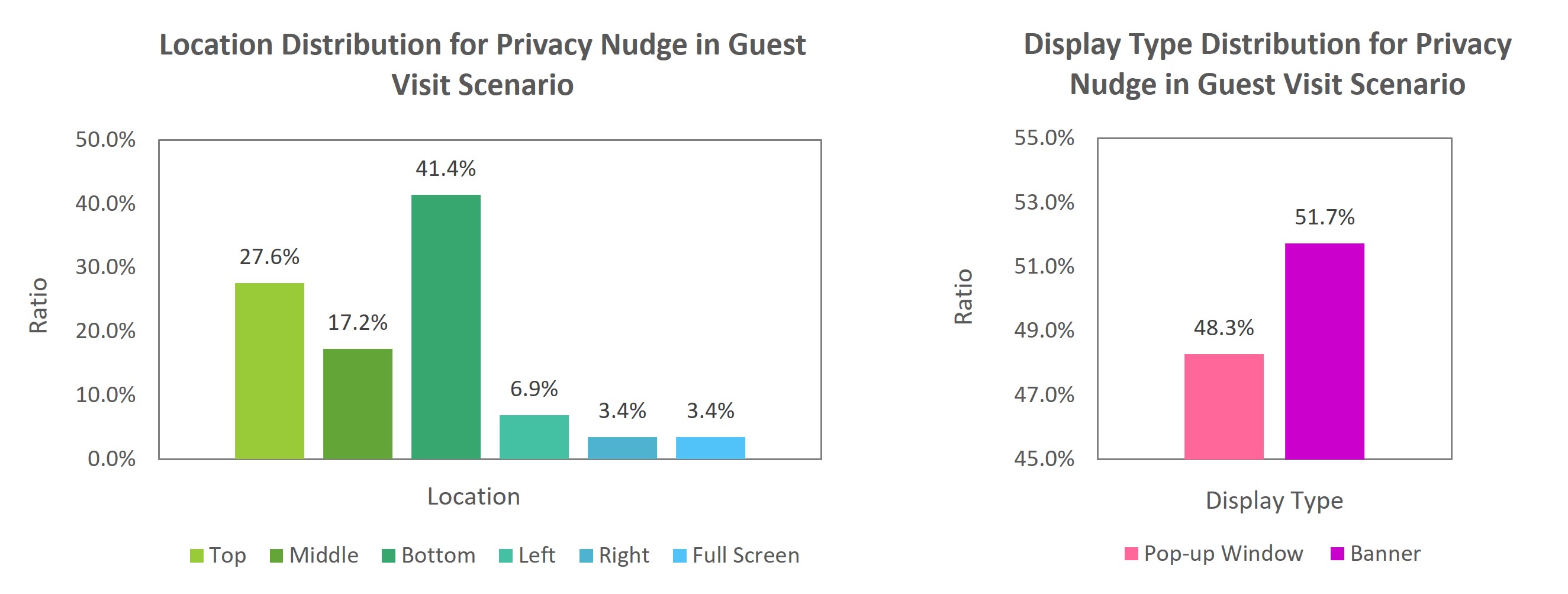}
  \caption{Privacy Nudge: Awareness Attribute Analysis in the Guest Visit Scenario}
  \label{nudge_guest}
\end{centering}
\end{figure}

\begin{figure}[h]
\begin{centering}
\includegraphics[width=0.8\textwidth]{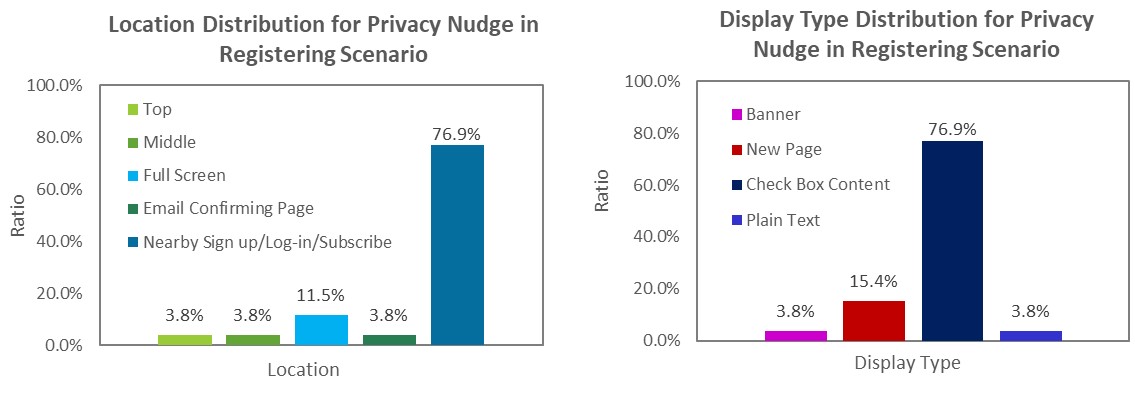}
  \caption{Privacy Nudge: Awareness Attribute Analysis in the Registering Scenario}
  \label{nudge_reg}
\end{centering}
\end{figure}

\begin{figure}[htb!]
\begin{centering}
\includegraphics[width=0.8\textwidth]{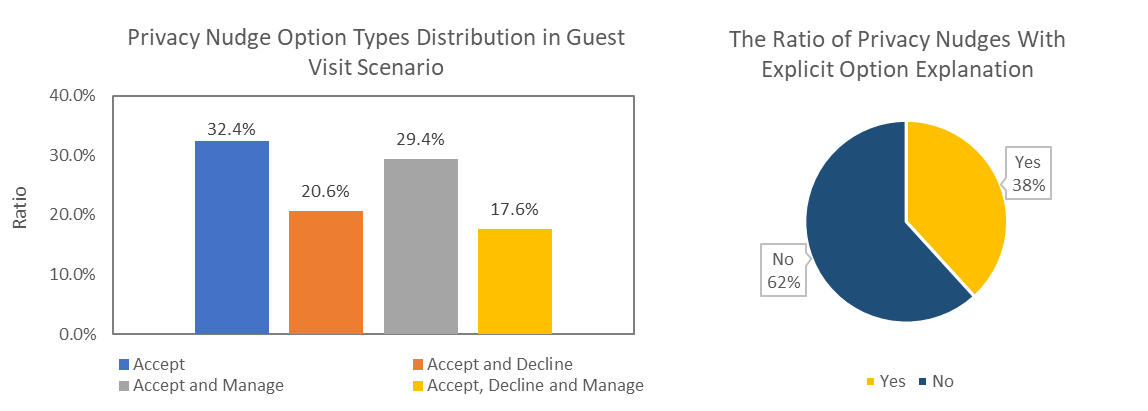}
  \caption{Privacy Nudge: Choice Attribute Analysis in the Guest Visit Scenario}
  \label{nudge_choice_guest}
\end{centering}
\end{figure}

\begin{figure}[htb!]
\begin{centering}
\includegraphics[width=0.8\textwidth]{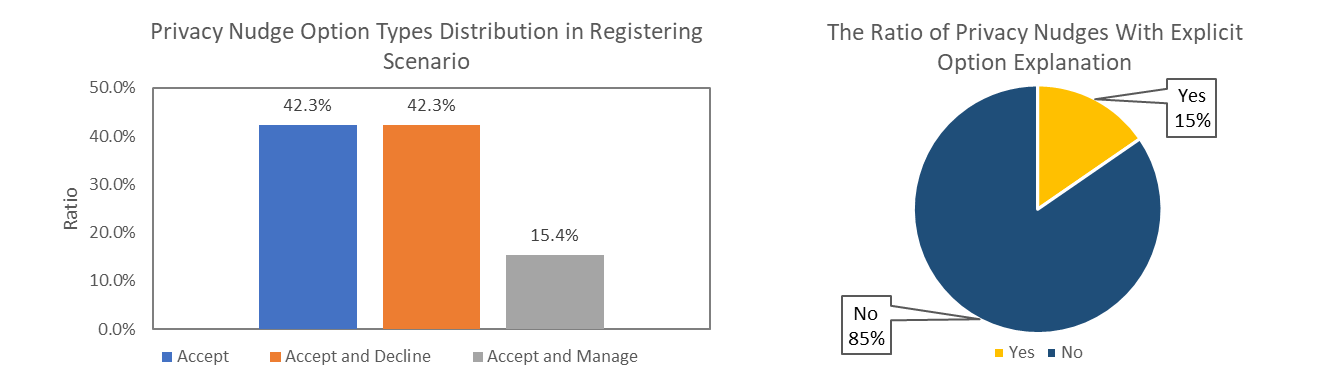}
  \caption{Privacy Nudge: Choice Attribute Analysis in the Registering Scenario}
  \label{nudge_choice_reg}
\end{centering}
\end{figure}

\begin{figure}[h]
\begin{centering}
\includegraphics[width=0.8\textwidth]{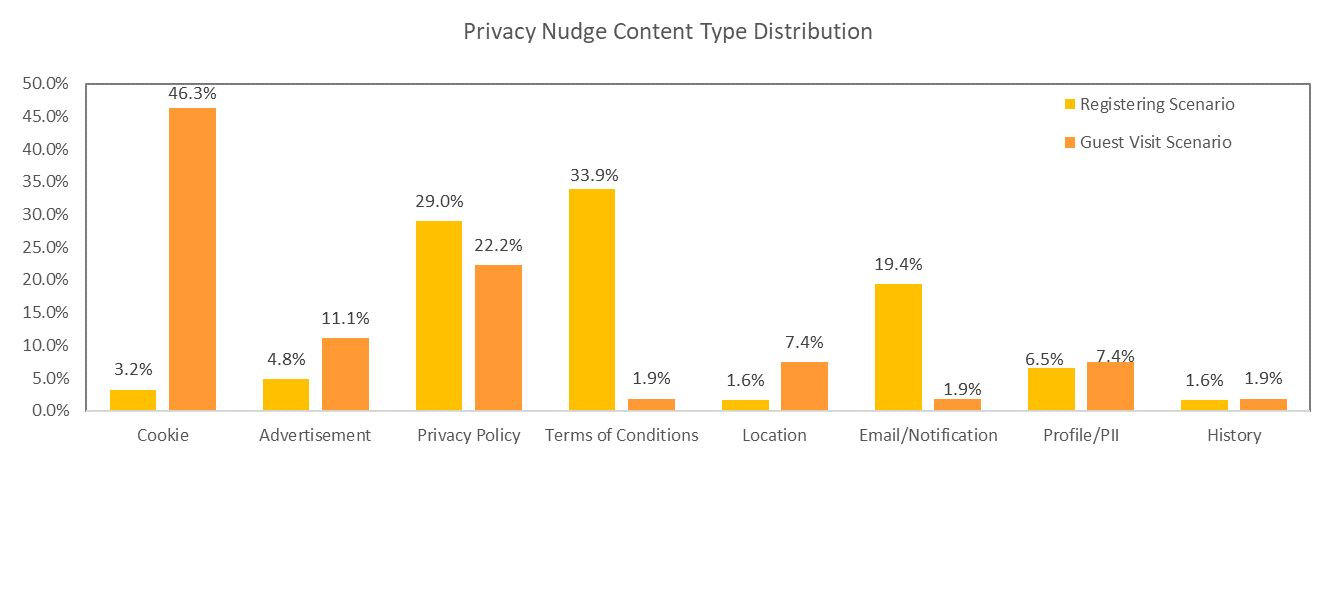}
  \caption{Privacy Nudge: Functionality Attribute Analysis in the Guest and Registering Scenarios}
  \label{nudge_function}
\end{centering}
\end{figure}

First, we present the data analysis related to the \textit{awareness} attribute of privacy nudge: location and display type (see Figure~\ref{nudge_guest} and Figure~\ref{nudge_reg}). Then we conduct the analysis of \textit{choice} attribute in Figure~\ref{nudge_choice_guest} and Figure~\ref{nudge_choice_reg}. Finally, we analyze the \textit{functionality} attribute of privacy nudges in Figure~\ref{nudge_function}.

\paragraph{Awareness} The location and type of privacy nudges are inconsistently displayed in the guest visit scenario, and most websites do not display a privacy nudge or display it inappropriately. Meanwhile, in the registering scenario, the privacy nudges are presented in a more consistent manner in terms of location and type, and most of the displayed privacy nudges are conspicuous.

There are 29 websites with the privacy nudge presented in the guest visit scenario. The nudges have six display locations and two display types (see Figure~\ref{nudge_guest}). The most common display locations in the screen for privacy nudges are the `Bottom' (41.4\%), `Top' (27.6\%), and `Middle' (17.2\%). The privacy nudges in the other three locations (`Left', `Right', and `Full Screen') occupy only 13.8\% in total. In terms of the distribution of display types, there is a nearly equal proportion of 'Pop-up Window' and 'Banner', with only a 3.4\% variation between the two. It is suggested that pop-ups perform better than banners when attracting the user's attention. 
According to the findings, the top banner outperforms the bottom banner; larger pop-ups or banners are more effective than smaller ones. Thus, ‘Pop-up Window', Banner' (located at the top only), and full-screen privacy nudges are considered 'effective' as they are easily noticeable by users. We count the number of these effective privacy nudges in the guest visit scenario and notice that 18 websites display privacy nudges in a salient way. That means 82 out of 100 websites do not have or only have subtle privacy nudges.

In the registering scenario, Figure~\ref{nudge_reg} shows that there are five display locations (i.e., `Top', `Middle' `Full Screen', `Email Confirming Page', and `Nearby sign up/log-in/subscribe') and four display types (i.e., `Banner', `New Page', `Check Box Content', and `Plain Text'). For the location, 76.9\% (20 out of 26) privacy nudges are displayed nearby sign up/log-in/subscribe buttons, and 11.5\% (3 out of 26) privacy nudges are shown with a full-screen web page. The privacy nudges located in other positions altogether only account for 11.5\%. For the display type, `Check Box Content' dominates the four display types, and `New Page' is the second most popular. The share of these two display types is 92.3\%. This suggests 92.3\% privacy nudges will not be ignored because a user needs to click the check box located nearby the sign up button or close the new page to continue the registering process.

\paragraph{Choice} The privacy nudges usually provide action options (i.e., choice) to a user for decision-making. For example, `Accept', `Decline', and `Manage'. Some privacy nudges only provide a single choice ('Accept'), while some provide the combination of `Accept' and another one or two options. The privacy nudges with only the `Accept' option disregard the user choice since the proposed option is mandatory. `Decline' give the right for users to reject sharing private information with websites. However, `Decline' may bring some limitations and impact users' browsing experience. The `Manage' option provides customized choices that maximize flexibility for users.
In the guest visit scenario, Figure~\ref{nudge_choice_guest} indicates that over half of the privacy nudges do not provide the 'Manage' option as well as an explicit explanation of the options. Similarly, Figure~\ref{nudge_choice_reg}) displays that only the minority (4) of privacy nudges provide a `Manage' option or explain the options explicitly in the registering scenario.

Specifically, in the guest visit scenario, 32.4\% of privacy nudges only provide the `Accept' option, which accounts for the highest proportion of all options. The share of the `Accept' and `Decline' combination is 20.6\%, and other combination groups with `Manage' together take up 47.0\%. On the other hand, most privacy nudges (61.8\%) do not provide a clear choice explanation. In the registering scenario, the share of the `Accept', `Accept and Decline', and `Accept and Manage' option is 42.3\%, 42.3\%, and 15.4\%, respectively. In addition, the privacy nudges with explicit option explanation account for only 15.4\%.

In summary, based on the choice attribute perspective, the majority of websites disregard user choice rights in both guest visits and registering scenarios. However, the privacy nudges in the guest visit scenario demonstrate better performance compared to those in the registering scenario, as they provide more user rights and clearer explanations of their options.

\paragraph{Functionality} The functionality of privacy nudges is various in the registering and guest visit scenarios. However, most privacy nudges only cover one or two of these functions such as cookie and privacy policy.
The privacy nudges push eight types of privacy content (i.e., functions): cookie, advertisement, privacy policy, terms of conditions, location, email/notification, personally identifiable information (PII), and history (see Figure~\ref{nudge_function}). 
We count the number of privacy nudges that have each type of privacy content to calculate the share proportion of every type of privacy content (see Table~\ref{table:function_number}). Some privacy nudges contain several types of privacy content. Therefore, these privacy nudges may be counted repeatedly. 

From Figure~\ref{nudge_function}, we can note that in the guest visit scenario,  cookies are the most commonly used privacy type, while in the registering scenario, terms of conditions have the highest share among the various privacy types. Location (7.4\%) and history (1.9\%) are the particular functions for the privacy nudges in the guest visit scenario. In addition, the privacy nudges with privacy policy take a significant proportion in type distribution, constituting similar proportions (22.2\% and 29.0\% respectively) in both scenarios. 

\begin{table}[h]
\caption{The Count for Websites Containing a Privacy Nudge with Different Number of Functions}
\label{table:function_number}
\footnotesize
\centering
\begin{tabular}{  lllllll }
\toprule
\textbf{Scenario}&\textbf{1}&\textbf{2}&\textbf{3}&\textbf{4}&\textbf{5}&\textbf{7} \\
\midrule
Guest Visit (29 total) &16 &8 &1 & &3 &1 \\
Registering (26 total) &16 &7 &2 & 1 & & \\
\bottomrule
\end{tabular}
\end{table}

From Table~\ref{table:function_number}, we notice that 55.2\% (16 out of 29) websites only cover one function in the guest visit scenario, whereas the ratio is 61.5\% (16 out of 26) in the registering scenario. Despite that there are five websites (guest) and three websites (registering) covering three or more functions, 82.8\% (guest) and 88.5\% (registering) of privacy nudges only contain two or fewer functions. For example, in the guest visit scenario, `abczdrowie.pl' is the only website presenting all seven functions in the privacy nudge. Only the website `fitbit.com' contains a maximum of four functions in the registering scenario.

\subsubsection{Privacy Notice}

\label{notice_analysis}
According to Table~\ref{table:usability}, there are only two usability attributes for the privacy notice control, that is \textit{awareness} and \textit{functionality}.

\paragraph{Awareness}
Similarly to the privacy nudge, the location and type of privacy notices are diverse. However, most websites do not contain a privacy notice or display it subtly in both visit scenarios.

\begin{figure}[h]
\begin{centering}
\includegraphics[width=0.9\textwidth]{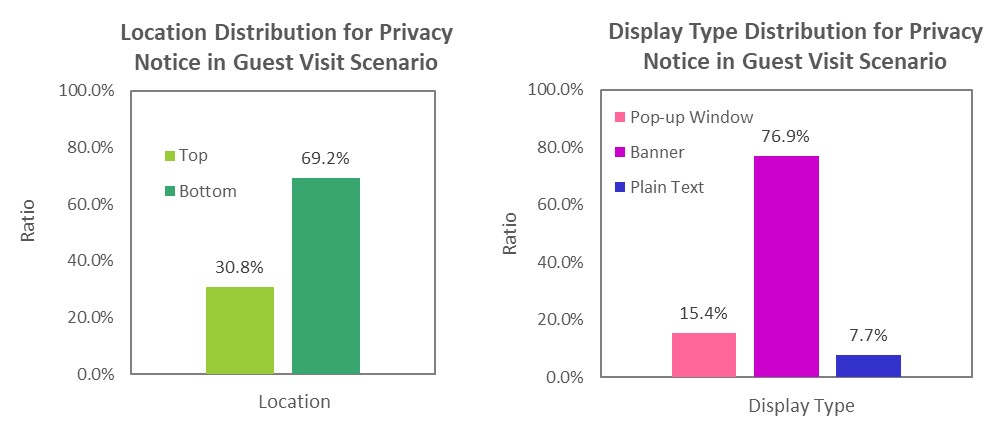}
  \caption{Privacy Notice: Awareness Attribute Analysis in the Guest Visit Scenario}
  \label{notice_guest}
\end{centering}
\end{figure}

In the guest visit scenario (see Figure~\ref{notice_guest}), the display location 'Bottom' and display type 'Banner', which are demonstrated to perform worse when attracting user attention, comprise 69.2\% and 76.9\%, respectively. More specifically, only 6 out of 100 websites display salient privacy notices with `Pop-up Window' or top `Banner'. The privacy notice and nudge utilized in the guest visit scenario can be perceived as intrusive since they may disrupt the browsing experience. However, if appropriately designed and presented, these mechanisms can effectively inform users about their rights regarding privacy control and increase their privacy awareness. They inform users of privacy-related information when users navigate the website for the first time. However, only 38 out of 100 websites contain either a privacy nudge or notice, and 21 websites amongst them prominently display a privacy nudge or notice.

\begin{figure}[h]
\begin{centering}
\includegraphics[width=0.9\textwidth]{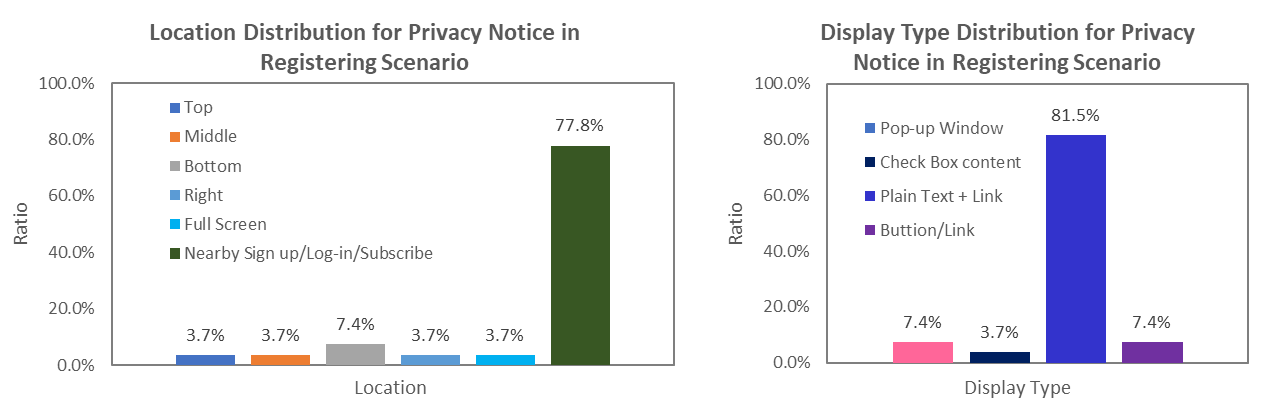}
  \caption{Privacy Notice: Awareness Attribute Analysis in the Registering Scenario}
  \label{notice_reg}
\end{centering}
\end{figure}

In registering scenario (see Figure~\ref{notice_reg}), the privacy notice can appear in five locations or be displayed on full screen. Like the privacy nudge, over three-quarters of the websites display privacy notices nearby sign up/log-in/subscribe buttons (first-ranking display location). The second-ranking display location is the `Bottom,' and its share is only about a tenth of the first-ranking location. Regarding the display type of the privacy notice, the dominant type is the `Plain Text and Link' (81.5\%). The plain text located nearby the sign up/log-in/subscribe buttons is not considered intrusive information as it does not disrupt the registration experience of users. 

Considering the privacy nudge and notice are complementary, we evaluate them together. In total, 39 out of 42 websites present either a privacy nudge or notice, and 29 websites are appropriately designed to present the privacy nudge or notice in an effective manner. The availability and usability (from an awareness perspective) of privacy nudge and notice in the registering scenario (92.9\% and 69.0\%) are much higher than those in the guest visit scenario (38.0\% and 21.0\%). 

\paragraph{Functionality}
Compared with privacy nudges, privacy notice contains fewer types of privacy content (i.e., functions), including three in the guest visit scenario and five in the registering scenario. The majority of privacy notices (100.0\% (guest) and 81.5\% (registering)) only cover one or two functions (see Table~\ref{table:function_number_notice}).

\begin{figure}[h]
\begin{centering}
\includegraphics[width=0.8\textwidth]{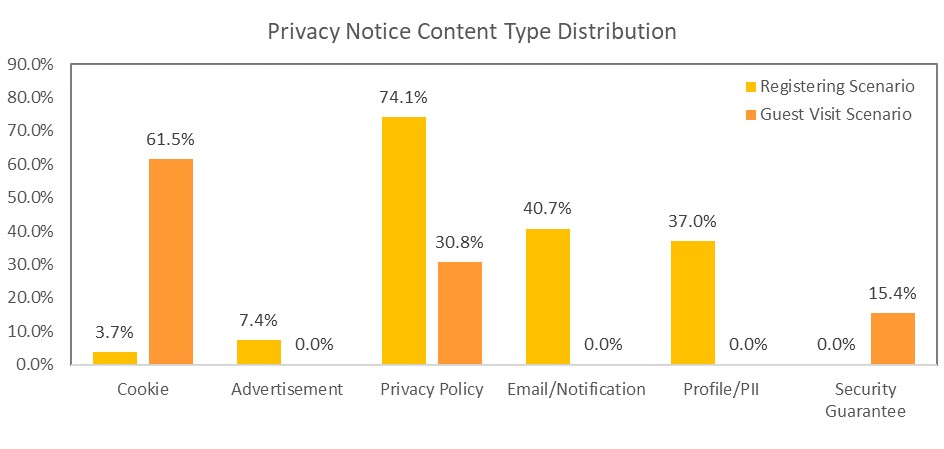}
  \caption{Privacy Notice: Functionality Attribute Analysis in the Guest and Registering Scenarios}
  \label{notice_function}
\end{centering}
\end{figure}

Figure~\ref{notice_function} represents the distribution of the privacy notice content type. The privacy notice in the guest visit scenario contains three types of functionality: cookie, privacy policy, and security guarantee. 61.5\% of privacy notices are related to cookie consent. In contrast to the guest visit scenario, the registering scenario includes three additional functions: advertisement, email/notification, and PII. We can observe that the privacy notice in different visit scenarios focuses on varying types of privacy content. For example, over half of the privacy notices in the registering scenario contain links to the privacy policy, but the share is only 30.8\% in the guest visit scenario. In addition, 40.7\%, 37.0\%, and 7.4\% of privacy notices inform the notification methods, the utilization of the PII, and the privacy content used for advertisement, but privacy notices in the guest visit scenario lack these types of privacy content. It is worth noting that, in the guest visit scenario, two government websites (i.e., `medicare.gov', `fda.gov') provide security guarantee information like `An official website of the United States government'. The security guarantee feature can alleviate users' privacy concerns, yet monitoring mechanisms are needed to prevent the abuse of such privacy notices.

\begin{table}[h]
\caption{The Count for Websites Containing a Privacy Notice With Different Number of Functions}
\label{table:function_number_notice}
\footnotesize
\centering
\begin{tabular}{  lllll }
\toprule
\textbf{Scenario}&\textbf{1}&\textbf{2}&\textbf{3}&\textbf{4} \\
\midrule
Guest Visit (13 total) &12 &1 & &  \\
Registering (27 total) &16 &6 &4 & 1\\
\bottomrule
\end{tabular}
\end{table}

\subsubsection{Privacy Policy}

\label{policy_analysis}
We analyze the four usability attributes (i.e., \textit{awareness, efficiency, comprehension} and \textit{choice}) for privacy policy complying with Table~\ref{table:usability}. It is unnecessary to focus on specific \textit{functionality}, as a privacy policy should encompass comprehensive information regarding a website's privacy. In the registering scenario, we only take into account the location feature of \textit{awareness}, as it may vary between the two scenarios. In terms of the \textit{'efficiency'} attribute, it is challenging to quantify the number of clicks to access the privacy policy since the registration process can have multiple variations, and the privacy policy link may appear at any point. Nevertheless, the remaining attributes of the privacy policy are consistent in both scenarios.

\paragraph{Awareness} In the guest visit scenario, the location of the privacy policy on the homepage is uniform, typically found in the footer section with either a button or a hyperlink display type. In the registering scenario, most privacy policies appear as links in privacy notices or nudges. 

\begin{figure}[htb]
\begin{centering}
\includegraphics[width=0.7\textwidth]{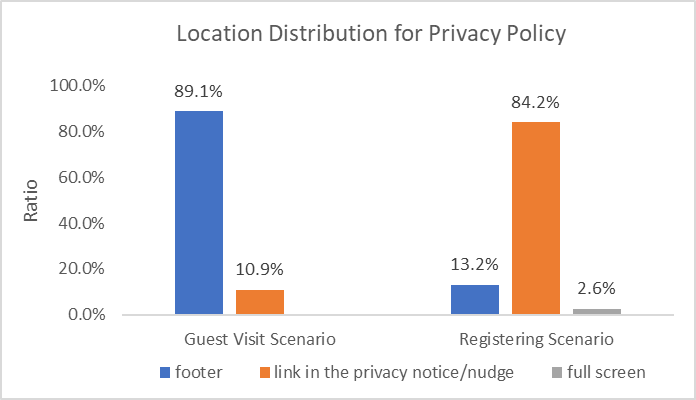}
  \caption{Privacy Policy: Awareness Attribute Analysis in the Guest Visit and Registering Scenarios}
  \label{policy_location}
\end{centering}
\end{figure}

For privacy policy analysis, we only focus on the display location instead of the display type, since we access the privacy policy either through a button or a link. Figure~\ref{policy_location} illustrates that 89.1\% of websites display the privacy policy button at the footer in the guest visit scenario. It is worth noting that the footer represents the final bottom portion of a web page and may contain multiple components. As such, the privacy policy may be situated within various components of the footer. In the registering scenario, the location of the privacy policy is not as uniform as in the guest visit scenario. Most privacy policies appear in the link of a privacy notice/nudge (84.2\%). One of these targeted websites, `altibbi.com', displays the privacy policy in a full-screen new page manner, which is considered a better practice to attract the user's attention than links in the privacy notice/nudge.

\paragraph{Efficiency} 
While two clicks are required to access the privacy policy of most websites through the privacy policy button or link, users may also access the privacy policy with only one click through a link in the privacy nudge/notice.

\begin{figure}[htb!]
\centering
\subfigure{\includegraphics[width=0.45\textwidth,trim={0 -1cm -1cm 0}]{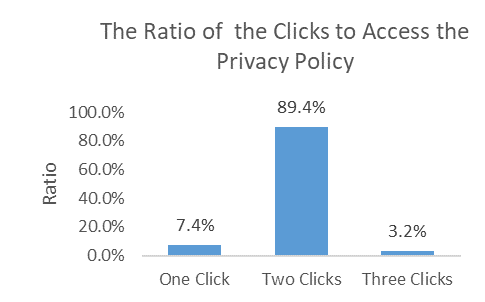}}
\subfigure{\includegraphics[width=0.45\textwidth,trim={-0.1cm -1cm -1cm 0}]{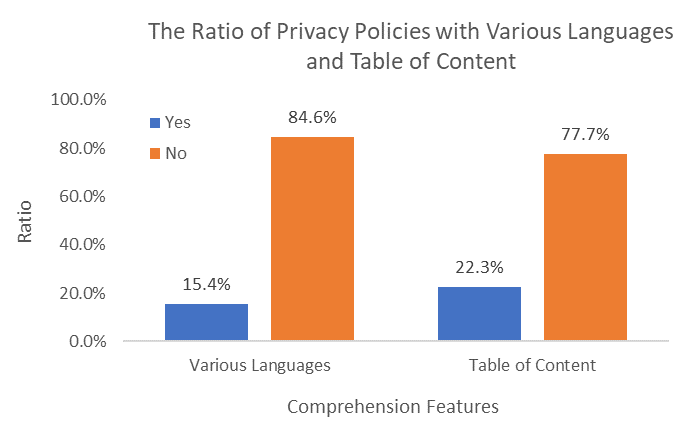}}
\caption{Privacy Policy: Efficiency and Comprehension Attribute Analysis}
\label{fig:policy}
\end{figure}

To represent the efficiency of visiting the privacy policy, we examine the number of clicks to access the privacy policy. 
While counting the clicks, we define the scrolling down action as one click. From Figure~\ref{policy_location}, we can see most privacy policy related buttons are at the footer; therefore, only two clicks (i.e., scrolling down and clicking on the button) are needed to access the privacy policy. Nonetheless, about 3.2\% of websites (see Figure~\ref{fig:policy} (left)) do not directly display the privacy policy after the button on the homepage is clicked, prompting users to perform an extra click. 
If the access to privacy policy appears in multiple locations such as in privacy nudge/notice (requiring one click) and in the footer (requiring minimum 2 clicks), we account for the lowest number of clicks required. In this case, 7.4\% of the websites present privacy policy access in privacy nudges/notices. Such location design is more efficient, as users only need to perform one click on the website to access the privacy policy.

\paragraph{Comprehension}
In this context, comprehension is not measured through 'readability', as previous research has illustrated the inadequate clarity of privacy policies in various domains. We utilize another two features to describe the ease of understanding the privacy policy, that is, whether the privacy policy is viewable in various languages and whether it contains a table of content to guide reading. The right sub-figure in Figure~\ref{fig:policy} demonstrated that the targeted websites perform poorly in terms of the comprehension attribute of the privacy policy. Over three-quarters of website privacy policies do not provide an alternative language version or a table of content to assist users in comprehending the document.

\paragraph{Choice}

\begin{figure}[htb]
\begin{centering}
\includegraphics[width=0.8\textwidth]{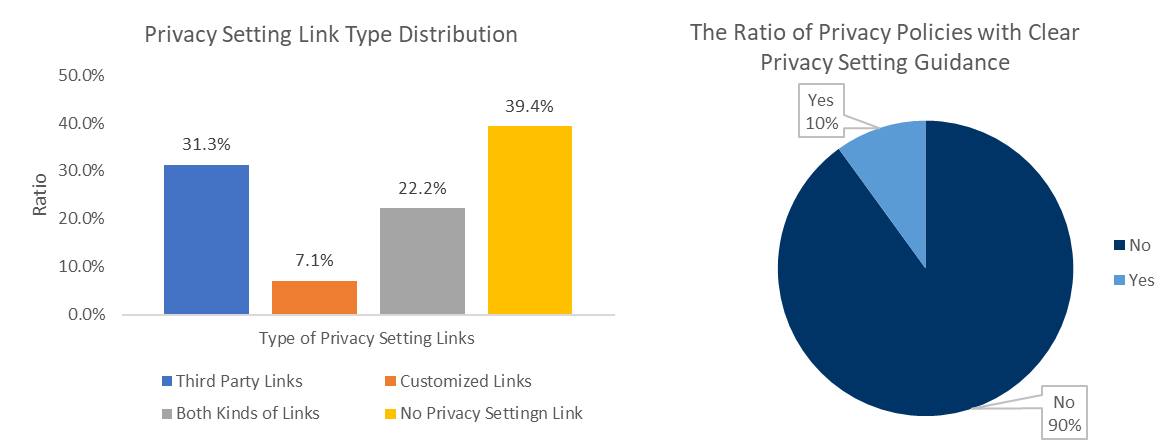}
  \caption{Privacy Policy: Choice Attribute Analysis}
  \label{policy_choice}
\end{centering}
\end{figure}

In the context of privacy policy analysis, the "type of option" refers to the type of privacy setting links that are available to users, it determines the ways in which users can exercise their privacy choices. In Figure~\ref{policy_choice} (left), we note that 39.4\% of privacy policies do not provide users any privacy setting links. Approximately one out of every three privacy policies exclusively present external links to third-party resources such as `All About Cookies', `Your Online Choices', `All About Do Not Track', `Network Advertising Initiative (NAI)' and links to privacy setting pages of Adobe and Google. Meanwhile, a significantly less proportion (7.1\%) of  privacy policies contain customized privacy setting links.

On the other hand, if the privacy policy provides clear guidance rather than meaningful privacy links for privacy settings, we can still treat the privacy policy as a good practice that respects users' `Choice' rights. Figure~\ref{policy_choice} (right) explains the comprehension attribute by evaluating the clarity of guidance. For the privacy policies that do not contain privacy setting links, we explore whether they provide clear privacy setting guidance. Only 4 out of the 40 aforementioned websites have explicit guidance. That means around 35.5\% of privacy policies completely forfeit users' privacy choice rights, providing none of the meaningful setting links or clear guidance.

\subsubsection{Privacy Settings}
\label{setting_analysis}
We analyze the usability of the privacy setting in each visit scenario based on multiple attributes. It should be highlighted that our analysis is limited to websites that have implemented privacy settings. Specifically, we analyzed a total of 34, 8, and 22 websites for the guest visit, registering, and log-in visit scenarios respectively.

\paragraph{Guest Visit Scenario}

\paragraph{Awareness} 

\begin{figure}[h]
\begin{centering}
\includegraphics[width=0.8\textwidth]{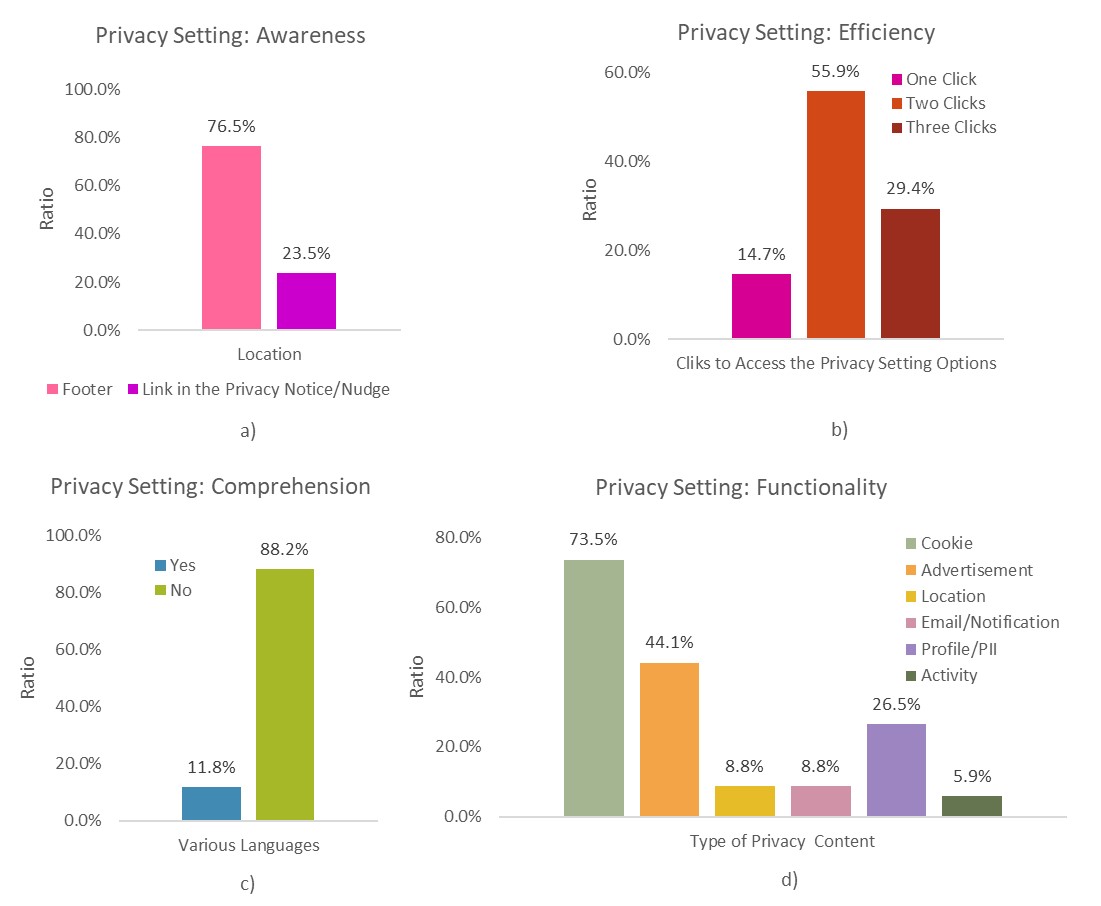}
  \caption{Usability Attribute Analysis for Privacy Setting in the Guest Visit Scenario}
  \label{setting_guest}
\end{centering}
\end{figure}

Figure~\ref{setting_guest} a) shows that similar to the privacy policy, the display locations of the privacy setting are relatively standardized. Such uniformity in display location is advantageous in facilitating users' ability to locate the privacy setting. About three-quarters of privacy setting buttons/links are located at the footer of a website, and the others (23.5\%) at the privacy nudge (7 out of 34)  or notice (1 out of 34). These privacy nudges serve as a conspicuous reminder to users regarding the availability of privacy controls on a website. This approach is highly effective in informing users about privacy settings, even if they possess limited knowledge about privacy protection.
Few websites (e.g., `shop-apotheke.com') provide the privacy setting button/link at both the footer and the privacy nudge, which is an optimal practice by combing both intrusive and voluntary methods. This ensures that the user can effectively perceive privacy controls and locate them effortlessly in a designated section whenever they require privacy protection.

\paragraph{Efficiency} We can access the privacy setting within three clicks on all targeted websites, with 14.7\% of them providing one-click access (see Figure~\ref{setting_guest} b)). All websites that with one-click access to the privacy setting present it through privacy nudges. It is worth noting that all these five websites are from the Europe region. On the other hand, 60.0\% of the websites with three-click access to the privacy setting (10 out of 34) are from North America. Therefore, we can deduce that in comparison to other regions, the design of privacy settings in Europe is superior. in the guest visit scenario from an \textit{efficiency} perspective.

\newpage
\paragraph{Comprehension}
Figure~\ref{setting_guest} c) shows that only about one-tenth of provide privacy settings in alternative languages.

For analyzing the privacy setting, we focus solely on the language aspect of the \textit{comprehension} attribute. In view of the restricted length of the privacy setting text, the inclusion of a table of contents is not obligatory. Even though 4 out of 34  websites provide multiple-language privacy settings, three of them are provided with a third-party link, `YourAdChoices', which has alternative language options. Another third-party link (`uptodate.com') provides alternative language options by redirecting websites to versions of different countries, like China (Chinese) and Brazil (Portuguese). That means ultimately none of the targeted websites offers a customized privacy setting interface with language options.

\paragraph{Functionality} Most privacy settings (73.5\%) in the guest visit scenario contain `Cookie' management options, and over half of the privacy settings only cover one function. 

Figure~\ref{setting_guest} d) summarizes the types of privacy content that privacy settings involve. The top three common types are the cookie, advertisement, and profile/PII. Other types of privacy content each account for less than 10.0\%. Table~\ref{table:function_number_setting} shows that 88.2\% (30 out of 34) privacy settings only cover one or two functions. For example, `menshealth.com' provides three functions through three buttons at the footer (`Manage Email Preferences',`Interest-Based Ads',`DO NOT SELL MY PERSONAL INFORMATION'). The website `familyandpets.com' provides four functions together with one toggle rather than separate toggles that correspond to each function. In this case, the best practices are the two Europe websites (`my-personaltrainer.it' and `medonet.pl'), which provide five functions in their unified privacy setting interfaces with independent toggles for each function.

\begin{table}[h]
\caption{The Count for Websites Containing a Privacy Setting With Different Number of Functions}
\label{table:function_number_setting}
\footnotesize
\centering
\begin{tabular}{  llllll }
\toprule
\textbf{Scenario}&\textbf{1}&\textbf{2}&\textbf{3}&\textbf{4}&\textbf{5} \\
\midrule
Guest Visit (34 total) &20 &10 &1 &1&2  \\
Registering (8 total) &6 &1 &1 & & \\
Log-in Visit (22 total) &15 &3 &3 & 1&\\
\bottomrule
\end{tabular}
\end{table}

\paragraph{Registering Scenario}

\paragraph{Awareness} Locating and the display pattern of different privacy settings can be a difficult task due to their diverse placement and format.

\begin{figure}[h]
\begin{centering}
\includegraphics[width=0.8\textwidth]{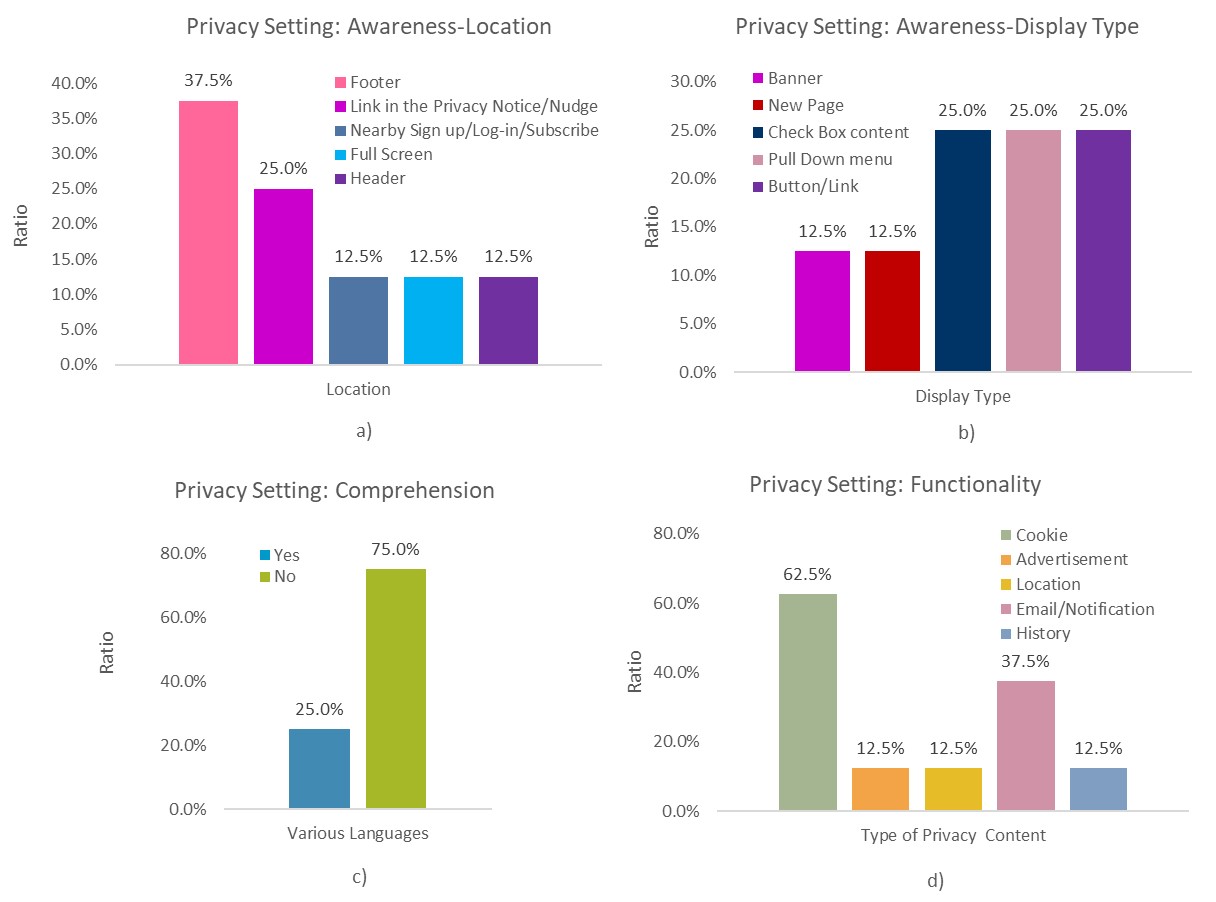}
  \caption{Usability Attribute Analysis for Privacy Setting in the Registering Scenario}
  \label{setting_registering}
\end{centering}
\end{figure}

In the registering scenario, we consider display location and type features (see Figure~\ref{setting_registering} a) and b)) for \textit{awareness} attribute. The display type of privacy setting during registration ranges diversely from banner, new page, check box content, and pull-down menu, which altogether accounts for 75.0\%. The privacy settings may appear at five display locations with five display types, generating a wide range of display combinations. No display location or type has a significant advantage. Therefore, it is difficult for users to form habits and build awareness of where the privacy setting will appear. 

\paragraph{Comprehension} Few websites (2 out of 8) provide multiple-language privacy settings (see Figure~\ref{setting_registering} c)). 

Similar to the guest visit scenario, these two websites do not offer language options on the privacy setting interface. For example, 'medscape.com' offers a selection of language at the beginning of the registration, instead of in the privacy setting (i.e., email notification preference) step. Once the registration process has commenced, users are unable to alter the language selection, and must repeat the registration procedure should they wish to do so.

\paragraph{Functionality} The majority of privacy settings only cover one function, and the most prevalent function is `Cookie'.

Table~\ref{table:function_number_setting} shows that only two websites cover more than one function. The other six websites provide a single function, constituting four of `Cookie' and two of `Email/Notification'. From Figure~\ref{setting_registering} d), we can see that the top two types of privacy content also are the cookie and email/notification. Each of the other three types of privacy content: advertisement, location, and history, occupies a small proportion of 12.5\%.

\newpage
\paragraph{Log-in Visit Scenario}
\paragraph{Awareness} Approximately 90.0\% of the account privacy settings are located at the header, which conforms to user account setup conventions.

\begin{figure}[hbt!]
\begin{centering}
\includegraphics[width=0.8\textwidth]{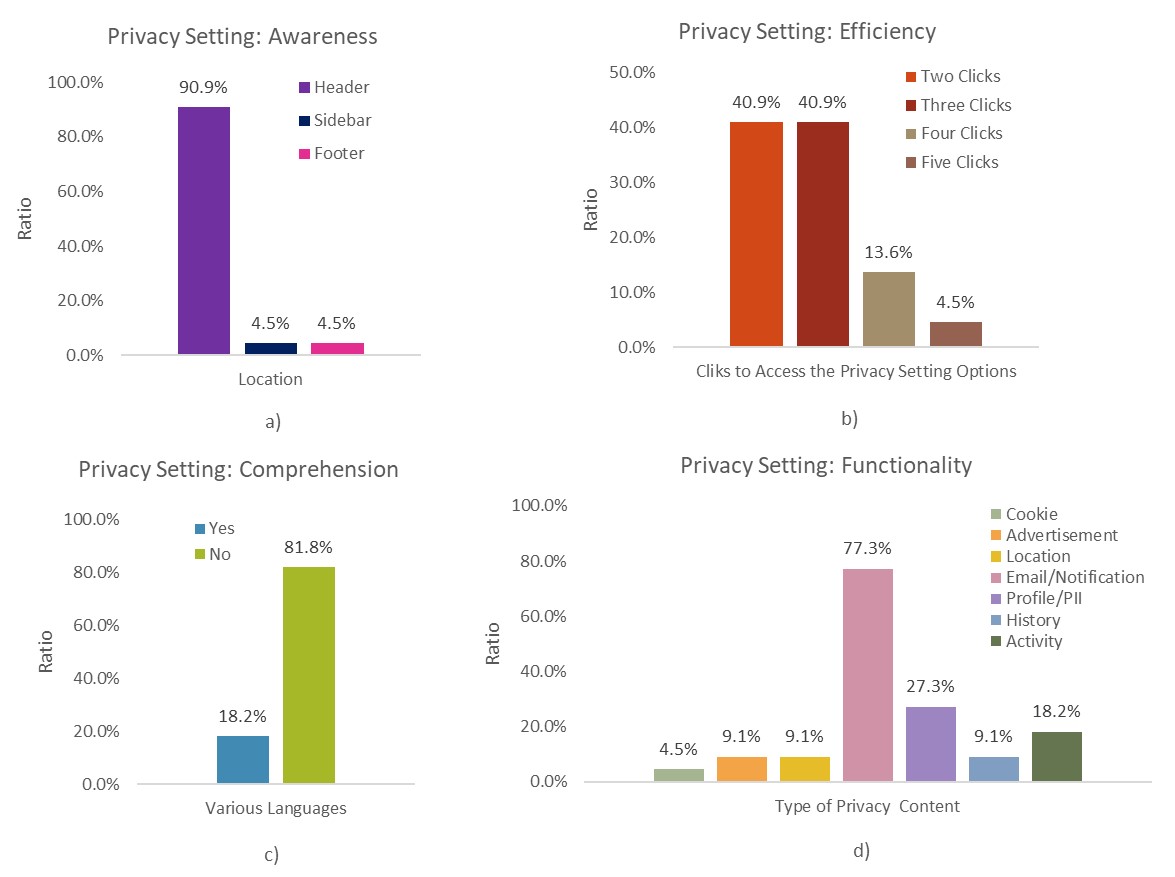}
  \caption{Usability Attribute Analysis for Account Privacy Setting in the Log-in Visit Scenario}
  \label{setting_login}
\end{centering}
\end{figure}

Figure~\ref{setting_login} a) shows that only 2 out of 22 (9.0\%) privacy settings appear at other locations rather than the header. For example, `amboss.com' displays the privacy setting with the `Account' button in the sidebar, which appears once users log in. Meanwhile, that sidebar can also be triggered by the click of the 3-line menu icon in the header.

\paragraph{Efficiency} Based on our findings, over 80.0\% (18 out of 22) of privacy settings can be accessed in an efficient way.

From Figure~\ref{setting_login} b), we observe that two-click and three-click access accounts for 40.9\% each. We consider the access within three clicks (included) efficient because the relationship between the clicks is straightforward. For example, the two-click access usually contains clicks on the header account setting icon and the privacy setting button (on the account setting page). Compared to the two-click access, the three-click access has an additional drop-down list after clicking the header account setting icon. 

On the other hand, about one-fifth of privacy settings can not be accessed efficiently. For example, the website `fitbit.com' needs five clicks to access the privacy setting, including clicks on the header account icon, the `My Dashboard' option in the menu, the setting icon in the header of the dashboard page, the `Setting' option in the menu, and the 'Privacy' option in the setting page. The account and setting icons may confuse users during the access process. It is possible for users to perceive that both icons lead to the same page, and consequently, disregard the second icon that links to the privacy setting page.

\paragraph{Comprehension} From Figure~\ref{setting_login} c), we conclude that only a few websites (4 out of 22) provide multi-language privacy settings. Similar to the other two visit scenarios, most websites do not display the language options on the privacy setting page. The website, `medscape.com', is a unique case since it is possible for users to customize the language setting on the privacy interface during the log-in visit.

\paragraph{Functionality} In the log-in visit scenario, the privacy setting tends to cover more functions, and the most prevalent type of content is `Email/Notification', instead of `Cookie' (see Figure~\ref{setting_login} d)) as in the other two visit scenarios. In line with another two visit scenarios, most privacy settings (15 out of 22) contain only one function (see Table~\ref{table:function_number_setting}).

There are seven types of privacy content (i.e., function): cookie, advertisement, location, email/notification, profile/PII, history, and activity. However, similar to the registering and guest visit scenarios, most privacy settings still cover merely one function. The top three functions are email/notification, profile/PII, and activity. Only one website (`doctolib.fr') covers the `Cookie' function. The website, `fitbit.com,' offers the widest range of functions and incorporates these functions into a privacy nudge during the registering scenario.

%% file: Sections/Discussion.tex
\section{Discussion}
\label{sec:discussion}

\subsection{Designing Privacy Controls}


In this section, we discuss how to improve the usability of each privacy control by complying with effective design principles.

\paragraph{Privacy Nudge and Notice}
From the 'awareness' perspective, the majority of websites (82.0\%  and 94.0\% respectively) lack clear privacy nudges and notices during guest visits. This suggests a need for designers to introduce more obvious privacy nudges or notices to increase users' awareness of privacy information. The use of intrusive display types, such as pop-up windows and banners placed consistently in the middle or top of the screen is recommended, with full-screen displays occupying the user's entire field of view being the most effective. Hence, we recommend websites use pop-up windows or full-screen new pages to deliver privacy nudges and notices in a conspicuous manner, instead of plainly displaying them as text near sign-up/log-in/subscribe buttons.
Moreover, due to the respect for users' freedom of choice, we suggest that privacy nudges should contain "Accept", "Reject", and "Manage" options, as well as an explanation for each option. However, survey results showed that only 10\% of websites in guest visit scenarios and none in registering scenarios satisfy this design requirement. Therefore, most websites need improvement in this aspect. From a functionality perspective, most privacy notices only cover one or two types of privacy content, such as cookies and privacy policies, and the functions covered depend on the data the website plans to collect.  The fifth clause of 999.305 b) in CCPA states that ‘business shall not collect categories
of personal information other than those disclosed in the notice at collection.’ Based on this design principle and
the analysis results (popularity of each function), we suggest that the privacy notice in guest visit scenarios should contain "Cookie", "Privacy Policy", "Location", and "Advertisement" functions, while the privacy notice in registering scenarios should contain  basic configurations like "Email/Notification" and "Profile/PII" functions in addition to the mandatory "Privacy Policy". Specific information like 'History' could also be added to the basic configuration according to a website's service.

\paragraph{Privacy Policy}

According to the results in Section~\ref{policy_analysis}, 
most privacy policies are located at the homepage footer in the guest visit scenario and displayed as links in privacy notices/nudges in the registering scenario. Policy. The two improvement suggestions for the display of a privacy policy are: firstly, providing a privacy policy link in a privacy nudge rather than a privacy notice that is easily ignored; secondly, displaying a privacy policy with a full-screen new page to attract users' attention to the maximum extent. Our study also suggests leveraging privacy nudge/notice functionality to access the privacy policy with only one click (i.e., provide privacy policy access in the privacy nudge/notice). Over three-quarters of privacy policies do not provide different language versions or a table of content, making it easy for users to get lost in the lengthy text. Therefore, the study recommended providing the privacy policy in multiple languages based on the potential user's first language and providing a table of content to help users locate the parts they are concerned about. Moreover, 36 out of 99 privacy policies did not provide privacy setting links nor clear privacy setting guidance. To fulfill the users' right to make decisions, we suggest the provision of customized privacy setting links where users can access all the privacy options with concise explanations (e.g., step-by-step), combined with third-party links with instructions for use. In general, implementing the recommended changes in the design of privacy policies has the potential to improve users' privacy protection and foster greater trust in the website.

\paragraph{Privacy Setting}

Based on our findings in Section~\ref{setting_analysis}, we provide suggestions to improve the display location, functionality, comprehension, and access efficiency of privacy settings. In the guest visit scenario, the privacy setting should be displayed in multiple formats, both as a button at the footer and as a link in the privacy nudge. To improve access efficiency, directly linking to the privacy setting from the privacy nudge is recommended. For the comprehension attribute, providing privacy settings in alternative languages is the better practice. For the registering scenario, we suggest providing check box content nearby the sign-up button or a full-screen privacy setting page after clicking the sign-up button. Such a design could ensure that privacy control is available before user information collection. Regarding the comprehension attribute, it is better to provide the privacy setting with multi-language versions or at least allow the privacy setting page to be automatically translated by the browser extension. Moreover, the privacy setting should cover ‘email/notification’ and ‘profile/PII’ functions besides ‘cookies’ and ‘location’. In the log-in visit scenario, it is recommended that the privacy setting should cover a full range of functions, including ‘cookies’, ‘advertisement’, ‘location’, ‘email/notification’, ‘profile/PII’, ‘history’, and ‘activity’. The location of the privacy setting should be consistent, unified, and easy to find, preferably in the account settings located in the header section. 

\subsection{Combining Privacy Controls}


Effective design of each privacy control is an important aspect of ensuring user privacy, but it is equally essential to understand the proper combination of these controls to achieve the maximum possible benefit for users. The privacy notice and privacy policy give users the right to know, focusing on delivering privacy-related information. On the other hand, the privacy nudge and privacy setting provide users the freedom of choice, which require users' consent or selection. According to the requirement of GDPR and CCPA, the most basic and core rights to users are the right to know and the right to choose (e.g., consent or opt-out). In addition, Art. 9 GDPR states that processing of data concerning health shall be prohibited unless the data subject has given explicit consent to the processing of those personal data. Therefore, health websites have higher requirements for privacy controls. We discuss the combination of privacy controls to deliver these rights to users in different visit scenarios.

\paragraph{Guest Visit Scenario} 
The privacy policy comprehensively describes a website's privacy practices and must be accessible in all visit scenarios. In the guest visit scenario, while users may not leave behind significant browsing traces or personally identifiable information on websites, the CCPA mandates that websites must provide users with prompt notice, either at or before the point of data collection. Consequently, more than just a privacy policy is necessary to fulfill the timely notice requirement. The minimum combination of privacy controls includes the privacy policy and the privacy nudge (intrusive style and contains customized options), or the privacy policy (contains setting options, links, or clear setting guidance) and the privacy nudge/notice (intrusive style). The best combination of privacy controls consists of the privacy policy, the privacy nudge (intrusive type and has customized options or privacy setting link), and the privacy setting.

\paragraph{Registering Scenario}

Under the CCPA notice requirement, websites shall provide new notice at or before data collection when they plan to collect additional categories of personal information. During the registration process, users are prompted to provide more personal information such as name, phone number, email, and address. Therefore, websites must provide notice during registration, using the same combination of privacy controls as in the guest visit scenario, except that intrusive privacy nudge/notice is not mandatory for the minimum configuration. Plain text or check box content placed near the sign-up button can also be used, as such an arrangement is clear and easily noticeable by users.

\paragraph{Log-in visit Scenario}
In the log-in visit scenario, users should have been aware of and consented to the privacy practices of this website. Therefore, the minimum configuration can include only the privacy policy and the best combination would be both privacy policy and privacy setting.
We observe from Figure~\ref{availability_scenario} that the availability of the privacy nudge, privacy notice, and privacy setting is relatively deficient compared with the privacy policy. We recalculate the availability of the combination of privacy controls in Table~\ref{table:availability_best_mini}.
The websites that can meet the minimum requirement in three visit scenarios account for 26.0\%, 57.1\%, and 69.0\%. The ratios of websites having the best configuration are much lower. In addition, we note that the regulatory compliance of the websites with the minimum configuration is over 80.0\%. From this finding, we can infer that regulations such as GDPR and CCPA provide guidance for designing websites, particularly in terms of ensuring compliance with relevant privacy requirements.

\begin{table} [h]
\caption{The Count for Websites With Different Privacy Control Configuration and Their Regulation Compliance}
\label{table:availability_best_mini}
\footnotesize
\centering
\begin{tabular}{  cccc }
\toprule
\textbf{Scenario}&\textbf{Minimum Configuration}&\textbf{Best Configuration}&\textbf{Regulatory Compliace}\\ 
\midrule
Guest Visit (100 total) &26 &13 & 80.8\% (21/26)\\ 
Registering (42 total) &24 &2 &87.5\% (21/24) \\
Log-in Visit (42 total) &29 &21 & 89.7\%(26/29) \\ 
\bottomrule
\end{tabular}
\end{table}

In summary, we recommend that website designers implement privacy controls following the best configuration to enhance user privacy, particularly in the guest visit scenario. It is also recommended that countries in the Asia-Pacific region establish or adopt privacy protection regulations similar to GDPR or CCPA to ensure the availability of privacy controls.

\subsection{Dark Patterns}

Dark patterns are increasingly being used to coerce unsuspecting users to accept less privacy protective options. During our data gathering process we came across several such practices followed by different websites. One of the most common dark patterns we came across is the cookie option selection window. When cookie choices are listed, by default, all cookies are enabled and the user has to individually turn off them one by one. Many websites do not provide the option to deselect all cookies with a single button. Another cookie related dark pattern we observed is the colour usage of the accept all button. Normally we associate the colour green with the more preferable option, and by having the accept all cookies in green a user can unintentionally click without reading the content properly. 

A similar by default opt-in behaviour can be seen during the registering scenario too. We identified that several websites have users opt-in to their newsletters and/or email subscriptions during the profile creation process. These options are mostly available in the form of checkboxes and placed closely to the terms and services agreement checkbox. This can confuse the user to think that all of these options should be ticked in order to create a user account resulting in an unwanted privacy setting.

\subsection{User Study}

Our current survey results are gathered by three researchers independently across three different time periods. This enabled us to capture the dynamic behaviour of privacy controls that are available in these websites. However, it is difficult to draw generalized conclusions about the usability of these websites perceived by the general public with this data. Another factor that should be taken into consideration is the domain knowledge of the researchers that currently gathered the data. Since they can be recognized as domain experts, their views and understandings of the privacy policies could be different from the average user. Therefore, to address these issues, we aim to conduct a user study covering a wide variety of user types. The same template we created will be used to gather data during the user study, with the same set of options being allowed under each question. However, a single user might not be able to evaluate all 100 websites in the survey. Hence, we can divide the 100 websites into groups that can be later combined by aggregating the results from several users to cover all 100 websites. The results of this study would provide us with a more detailed and accurate description of the level of usability realized by the general public.


%% file: Sections/Conclusion.tex
\section{Conclusion}
\label{sec:conclusion}

We gathered information related to privacy controls of the top 100 most visited health websites using a analysis template designed by us. This template covers all four privacy controls; privacy nudges, privacy notices, privacy policies, and privacy settings. With the implementation of privacy regulations such as GDPR and CCPA, websites are offering more control and transparency over the privacy of the users visiting the website. However, the results from our survey showed that even though there are some forms of privacy controls made readily available for users, most of them lack the usability aspect. The unstructured display of privacy notices and nudges, the absence of clear instructions under privacy settings, and the limited language options provided for privacy policies are few of the issues we have identified. To this end, we suggested several privacy control designs and combined guidelines to improve the shortcomings present in these websites.